\documentclass[traditabstract]{aa}
\usepackage{psfrag,graphicx}
\usepackage{txfonts}
\usepackage{natbib}
\usepackage{ulem}
\usepackage{color}
\usepackage{xspace}
\usepackage{enumitem}
\usepackage{lipsum}
\usepackage{relsize}
\usepackage{tablefootnote}

\def \cmcube{\,cm$^{-3}$\xspace}

\DeclareMathAlphabet{\mathpzc}{OT1}{pzc}{m}{it}

\begin{document}
\bibliographystyle{aa}

\title{Parameterizing the interstellar dust temperature}
\subtitle{}
\titlerunning{Parameterizing the dust temperature}
\author{S. Hocuk \inst{1} \and L. Sz\H{u}cs \inst{1} \and P. Caselli \inst{1} \and S. Cazaux \inst{2,3} \and M. Spaans \inst{2} \and G. B. Esplugues \inst{1,2}
       }
\institute{
  $^1$ Max-Planck-Institut f\"{u}r extraterrestrische Physik, Giessenbachstrasse 1, 85748 Garching, Germany\\
  $^2$ Kapteyn Astronomical Institute, University of Groningen, P. O. Box 800, 9700 AV Groningen, Netherlands \\
  $^3$ Leiden Observatory, Leiden University, P.O. Box 9513, NL 2300 RA Leiden, The Netherlands \\
  \email{seyit@mpe.mpg.de, laszlo.szucs@mpe.mpg.de}
}
\authorrunning{Hocuk, Sz\H{u}cs, Caselli, Cazaux, Spaans, Esplugues}
\date{Received \today}

\abstract
{The temperature of interstellar dust particles is of great importance to astronomers. It plays a crucial role in the thermodynamics of interstellar clouds, because of the gas-dust collisional coupling. It is also a key parameter in astrochemical studies that governs the rate at which molecules form on dust. In 3D (magneto)hydrodynamic simulations often a simple expression for the dust temperature is adopted, because of computational constraints, while astrochemical modelers tend to keep the dust temperature constant over a large range of parameter space. Our aim is to provide an easy-to-use parametric expression for the dust temperature as a function of visual extinction ($A_{\rm V}$) and to shed light on the critical dependencies of the dust temperature on the grain composition. We obtain an expression for the dust temperature by semi-analytically solving the dust thermal balance for different types of grains and compare to a collection of recent observational measurements. We also explore the effect of ices on the dust temperature. Our results show that a mixed carbonaceous-silicate type dust with a high carbon volume fraction matches the observations best. We find that ice formation allows the dust to be warmer by up to 15\% at high optical depths ($A_{\rm V}> 20$\,mag) in the interstellar medium. Our parametric expression for the dust temperature is presented as $T_{\rm d} = \left[ 11 + 5.7\times \tanh\bigl( 0.61 - \log_{10}(A_{\rm V})\bigr) \right] \, \chi_{\rm uv}^{1/5.9}$, where $\chi_{\rm uv}$ is in units of the \cite{1978ApJS...36..595D} UV field.
}

\maketitle

\section{Introduction}
\label{sec:introduction}
Dust chemistry plays an important role during the evolution of interstellar clouds. The presence of dust is ubiquitous in the interstellar medium (ISM) and it is an important constituent of the Galaxy. Having a mass of only about 0.7\% of the gas \citep{2014Natur.505..186F}, these microscopic particles greatly impact the chemistry and thermodynamics of gaseous clouds \citep{1978ApJS...37....1G, 2013ApJ...766..103D, 2014MNRAS.438L..56H}, along with dominating the continuum opacity. Gas-phase species can use grain surfaces as a third body to form more complex molecules, thereby catalyzing reactions which may otherwise be too slow to be significant \citep[e.g.,][]{1963ApJ...138..393G, 2005JPhCS...6..155C, 2008ApJ...682..283G, 2012MNRAS.424.2961G, 2015MNRAS.447.4004R}. Atoms and molecules can, on the other hand, also be depleted from the gas phase when the dust temperature is too cold for species to overcome the thermal desorption energy \citep[e.g.,][]{1984MNRAS.209..955J, 2013A&A...560A..41L}. In this way, the dust temperature crucially controls whether gas-phase species freeze out onto dust, or are enriched from the chemistry occurring on dust grains \citep{2002ApJ...569..815T, 2006A&A...457..927G, 2016MNRAS.456.2586H}.

The dust temperature is also an important parameter in chemical reaction rates. An exponential dependence on the dust temperature lies at the heart of most surface reactions. At cold 10\,K temperatures, for example, the difference of a single Kelvin can imply a variation of the reaction rates by orders of magnitude. Thus, a precise knowledge of the dust temperature is imperative when performing rate calculations. Fortunately, calculated abundances seem to be more sensitive to the relative reaction rates between those which compete with each other rather than the absolute reaction rates \citep[see e.g.,][]{1982A&A...114..245T, 2007A&A...469..973C, 2016A&A...585A..55C}. Nonetheless, a study on the dependencies of the dust temperature is highly desirable.

A third and fundamental importance of the dust temperature follows from its impact on the gas temperature. The gas-dust collisional coupling is the single most important heat transfer mechanism for gas at number densities above a few\,$\times$\,10$^4$\,cm$^{-3}$ for typical Galactic conditions \citep{1991ApJ...377..192H, 2006ApJ...652..902S}. This process dominates the gas cooling as long as the dust temperature is lower than the gas temperature. At densities of roughly $>$10$^6$\,cm$^{-3}$, the temperatures of the two phases are irrevocably linked, such that the gas temperature is essentially set by the dust temperature. In regions with density below $\sim$10$^{4.5}$\,cm$^{-3}$, where line emission regulates the gas temperature, the temperature of the dust still plays a role. Here, dust can influence the gas temperature because the molecular abundances of species such as CO and H$_2$ (at T\,$>$\,100\,K), which depend on surface chemistry, control the amount of ro-vibrational line cooling \citep[see e.g.,][]{2016MNRAS.456.2586H}. 

In this work, we have derived the dust temperature semi-analytically for various types of grain material through solving the energy balance. We compare our results to a collection of observed dust temperatures with the \textit{Herschel Space Observatory}. In this way, by constraining our calculations with the observed dust temperatures, our study sheds light on the composition of dust in the ISM. The dust temperature is also explored for the presence of ices on dust surfaces. To substantiate our method, we test our semi-analytical solutions against a numerical one, i.e., radiative transfer calculations.

In Sect.\,\ref{sec:obs}, we report on a collection of observations of the interstellar dust temperature from the literature. In Sect.\,\ref{sec:theory}, we describe the analytical method that we use in order to derive the dust temperature, where we discuss the parameter details in Sect.\,\ref{sec:parameters}. We present our semi-analytical solutions for the dust temperature for various grain materials as well as for ice coated grains in Sect.\,\ref{sec:self}. In Sect.\,\ref{sec:radmc-3d}, we compare our semi-analytical solutions against computations with the Monte Carlo radiative transfer code \textsc{radmc-3d}\footnote{http://www.ita.uni-heidelberg.de/$\sim$dullemond/software/radmc-3d} \citep{Dullemond}. We then introduce our parametric expression for the dust temperature in Sect.\,\ref{sec:expression}. In Sect.\,\ref{sec:combine}, we discuss our theoretical solutions with respect to the observed dust temperatures. Finally, in Sect.\,\ref{sec:conclusion}, we summarize our conclusions.

\section{Observed dust temperatures}
\label{sec:obs}
We first looked at observations reported in the literature to check if a simple expression for the dust temperature as a function of $A_{\rm V}$ can be found. We show that this is not practical because one is probing different environments and different conditions among the various sources included and because of a missing general understanding of the physics. We will use these observations at a later stage to derive our parametric formalism from the inferred dust properties (in Sect.\,\ref{sec:combine}).

\subsection{Herschel observations}
Recent observations with the \textit{Herschel} space telescope provided a large number of measurements of the dust temperature in various sources. These sources consist of dense filaments, clumps, starless and prestellar cores, and protostellar cores. We select eight independent studies that report the dust temperature for a variety of sources and adopt their values, but exclude the protostellar cores from our selection as these have fundamentally different environments due to protostellar feedback. We present a compilation of the published dust temperatures at the specified column densities, $N_{\rm H}$, or $A_{\rm V}$. To stay within the same units, we convert $N_{\rm H}$ to $A_{\rm V}$ using the conversion factor $2.2\times10^{21}$ cm$^{-2}$ mag$^{-1}$ \citep{2009MNRAS.400.2050G, 2015ApJ...809...66V}.

\begin{figure}
\includegraphics[scale=0.51]{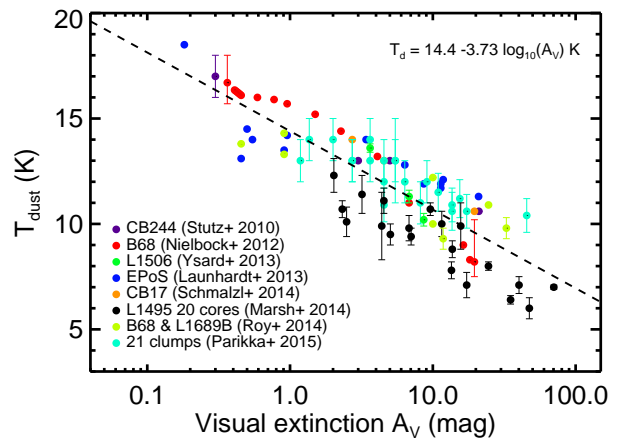}
\caption{Observed dust temperatures from eight independent studies. The dust temperature is plotted as a function of visual extinction. A least squares semi-log linear line is fit through the data as given by the dashed black line. }
\label{fig:td3}
\end{figure}

Fig.\,\ref{fig:td3} displays the collection of observationally obtained dust temperatures, with the references given in the legend of the figure.
We have drawn the best fit semi-log linear line through the data points, but the functional form is arbitrary. The fitting function we obtain is $T_{\rm d}(A_{\rm V}) = 14.4-3.73\log_{10}(A_{\rm V})$\,K. However, there is no physical basis for this behavior and such a fit cannot show the dependence in the radiation field. Furthermore, by using this fit, one is not able to extrapolate with certainty beyond the bounds of the observed range ($A_{\rm V} \simeq 0.2-70$\,mag).

The error bars are considered wherever they are available, though, only for the dust temperatures. The error bars for $N_{\rm H}$ or $A_{\rm V}$ are often not reported and, if given, can have large uncertainties. For example, in the case of the 21 cold clumps study by \cite{2015A&A...577A..69P}, two methods for calculating $N$(H$_2$) are given, that is, from dust continuum and molecular lines, which diverge greatly in some cases and thereby influence the results. We adopted the one that is recommended (dust continuum) by these authors. For the study of starless and prestellar cores in L1495 of Taurus by \cite{2014MNRAS.439.3683M} we recovered the column densities by computing this ourselves using the provided number densities, radii, and the Plummer-like density profile.

From the 14 low-mass molecular cloud cores \citep[EPoS project,][]{2013A&A...551A..98L}, we took the 7 starless cores and excluded the protostellar cores. Also from the bok globule CB244 \citep{2010A&A...518L..87S}, we only considered the starless core measurement. The only filament in our collection is the dense filament of the Taurus molecular complex L1506 \citep{2013A&A...559A.133Y}, which appears to have a density of $n_{\rm H} > 10^3$\cmcube, where a 3D radiative transfer model is used for estimating the emission and extinction of the dense filament.

In the studies of the isolated starless core B68 \citep{2012A&A...547A..11N}, the star-forming core CB17 \citep{2014A&A...569A...7S}, the dense cores in the L1495 cloud of the Taurus star-forming region \citep{2014MNRAS.439.3683M}, and the starless cores B68 and L1689B \citep{2014A&A...562A.138R} various techniques have been used to remove the line-of-sight (LOS) contamination, always resulting in a lower dust temperature (by about $0-4$\,K) than the ones obtained from dust spectral energy distributions (SED) only\footnote{When simply using SED fitting, the average dust temperature along the sight line is obtained.}. The used LOS correction techniques are, in the order of the above listed sources, an employed ray-tracing model, a modified black body technique together with a ray-tracing technique, a radiative transfer model (\textsc{corefit/modust}), and an inverse-Abel transform-based technique.

Despite such mixed origins in Fig.\,\ref{fig:td3}, i.e., with and without LOS corrections, the difference is not directly obvious from the plot, except that data points corrected for LOS effects generally have a lower temperature toward higher $A_{\rm V}$. This can be perceived by looking at the red and black points.

\subsection{Environmental differences}
The external radiation field strength in many of these sources is not known. In units of the Habing field \citep{1968BAN....19..421H}, the ones that are known have the best estimates of $G_0 = 1$ for L1506 and B68 \citep{2013A&A...559A.133Y, 2014A&A...562A.138R}, $G_0$ = 0.18 to 1.18 for the dense cores in L1495 \citep[][their standard $\chi_{\rm ISRF}$ corresponds to $G_0=1.31$]{2014MNRAS.439.3683M}, and $G_0 \approx 2$ for L1689B \citep{2016A&A...593A...6S}, which was initially (over)estimated to be $G_0 \approx 10$ \citep{2014A&A...562A.138R}. For the cores CB244, B68, and CB17, \cite{2016A&A...592A..61L} estimate an enhancement of factor 2.5, 2.2, and 3.0, respectively, relative to their interstellar radiation field (ISRF). The thermal dust temperature in the diffuse medium surrounding the remaining objects ranges between $\sim 16$ to 20\,K. These values are typical for the Milky Way diffuse ISM and the ISRF is thus assumed to be close to the standard Galactic radiation field ($G_0 \approx 1-2$).
Other differences, such as, number density, dust-to-gas mass ratio, type and size of grains, turbulence, and magnetic field are all important factors not taken into account. For these reasons, we targeted similar types of environments, the starless cold dense regions, where the above mentioned conditions are not expected to vary greatly. In spite of the unknown intrinsic differences, we still proceeded to overlay the various observations in a single plot (i.e., Fig.\,\ref{fig:td3}) to give us a general indication about the dust temperature in such environments.

\section{Solving for the dust temperature}
\label{sec:theory}
A simple way of obtaining the dust temperature is by solving the dust thermal balance for equilibrium. The underlying assumption is that an equilibrium is quickly reached and maintained. The primary heating and cooling processes are fast enough, on the order of $10^{-5}$ seconds (radiative cooling) to minutes and hours \citep[heating by interstellar photons, e.g.,][]{2003ARA&A..41..241D}, to justify this approach. Presuming that the main heating of dust is caused by the ISRF and that the primary cooling comes from the isotropic modified\footnote{Here we mean lower than unity emissivity, but with $\nu$ dependence.} black body emission, the energy balance for a single grain can be set up as follows \citep[cf.][]{2008ipid.book.....K}
\begin{equation}
4\pi a^2 \int^{\infty}_{0} Q_{\nu} B_{\nu} (T_{\rm d}) \, d\nu = 2\pi a^2 \int^{\infty}_{0} Q_{\nu} J_{\nu} D_{\nu} (A_{\rm V}) \, d\nu.
\label{eq:thbalance}
\end{equation}
Here, $Q_{\nu}$ is the absorption efficiency (Sect.\,\ref{sec:Q}) that depends on $a$, the grain radius, $T_{\rm d}$ is the dust temperature, $B_{\nu}$ is the Planck function, $J_{\nu}$ is the ISRF flux (Sect.\,\ref{sec:isrf}), and $D_{\nu}$($A_{\rm V}$) is the attenuation factor (Sect.\,\ref{sec:attenuation}), with $A_{\rm V}$ the visual extinction. In this work, we assume a geometry that is spherically symmetric and that the interstellar radiation is coming from all directions, but that the cloud is large enough to shield the radiation from one side. Hence, we take {\boldmath $2\pi$} for the right-hand side. Our choice is best described by a semi-infinite slab. This represents the edges of a cloud very well, whereas the center of a cloud should tend to $4\pi$. 
Assuming that a medium is in radiative equilibrium, for a distribution of dust grains, one may apply a second integral over grain sizes in Eq.\,\ref{eq:thbalance}. The integral equation becomes independent of $a$ if a fixed size is adopted. In this work, due to the complex and evolving nature of dust grains, we present our solutions for the canonical size of $a=0.1\,\mu$m \citep[e.g.,][]{2003pid..book.....K}. 

This relatively simple concept is often adopted to obtain the dust temperature. This usually involves assumptions for certain aspects of the calculation (i.e., $Q_{\nu}$, $J_{\nu}$, or $D_{\nu}$) or is simplified by limiting the solution to a desired range, which we discuss in the next section. Equilibrium solutions are, of course, always time independent and tend to consider simple geometries, like a slab or a sphere. The benefit is that the calculation is fast and stable, while the solutions are considered to be satisfactory.

It is advisable to note that for small dust grains ($a \lesssim 50$\,\AA), the equilibrium solution will not hold, since there will be large temperature fluctuations following single-photon heating events \citep{2001ApJ...551..807D}. Although the mass in small grains is low, a substantial fraction of the emission from diffuse clouds may be coming from them \citep{2001ApJ...551..807D, 2001ApJ...554..778L}. Non-equilibrium solutions should therefore be used for a better treatment of very small grains in diffuse regions. 

One can expand the energy balance equation by adding more heating and cooling terms. One factor that may be important for the heating of dust grains are cosmic rays (CR). Upon hitting a dust grain or a molecule, cosmic ray particles can either heat the grain locally, i.e., impulsive spot heating \citep{1985A&A...144..147L, 2015ApJ...805...59I}, or globally, e.g., due to secondary UV photons generated following H$_2$ fluorescence in the Lyman and Werner bands. CRs are insensitive to a gas column density of up to $N_{\rm H} \gtrsim 10^{23} \rm cm^{-2}$ \citep{2009A&A...501..619P, 2015ApJ...800...40I} and have an attenuation length of $N_{\rm H} \simeq 6\times10^{25} \rm cm^{-2}$ \citep{1981PASJ...33..617U}. This renders the energetic secondary UV photons nearly independent on the cloud optical depth, which results in a constant contribution to the right-hand side of the energy balance equation. For heating the dust grains by secondary UV photons, adopting a CR induced UV photon (CRUV) flux of $F_{\rm UV} = 2\times10^{4} \rm \,s^{-1}\,cm^{-2}$ \citep[the `low' proton model of][]{2015ApJ...805...59I} and a photon energy of 13\,eV per CRUV, the total intensity becomes $I_{\rm CR} = 4.2\times10^{-7} \rm \,erg \,s^{-1} \,cm^{-2} \,sr^{-1}$. From these estimations, we can already report that the CR impact on the dust temperature for a standard Milky Way ionization rate, i.e., $\zeta_{H_2} = 5\times10^{-17}$\,s$^{-1}$, turns out to be minimal. We find that due to CRs the dust temperature increases by $\Delta T_{\rm d} \lesssim$\,0.1\,K. Higher CR rates have been reported \citep[e.g.,][]{2015ApJ...800...40I} and some CR attenuation is expected \citep[e.g.,][]{2015ApJ...805...59I}. The impact of higher cosmic ray rates for the considered simplified case (i.e., without attenuation) is explored in appendix\,\ref{app:sec:crdr}.

\section{Parameter details}
\label{sec:parameters}
\subsection{$J_{\nu}$, the ISRF}
\label{sec:isrf}
Each parameter in the integral of Eq.\,\ref{eq:thbalance} is a function of frequency that goes from 0 to infinity. In actuality, however, this limit is set by the ISRF, which affects the other parameters. Typically, the Galactic ISRF covers the wavelength regime between microwave (3000\,$\mu$m) and far-ultraviolet (FUV, 0.1\,$\mu$m), with contributions from stars (OB stars and late spectral classes), dust (cold, warm, and hot), and the cosmic microwave background (CMB). Starlight dominates the emission between the FUV and the near-infrared (NIR), while dust mainly emits reprocessed starlight between the NIR and the far-infrared (FIR). Beyond this range, there is little emission to be significant for the dust temperature. The shorter wavelengths are more important toward lower $A_{\rm V}$, while the longer wavelengths become more relevant at higher optical depths.

\begin{figure*}
\centering
\includegraphics[scale=0.51]{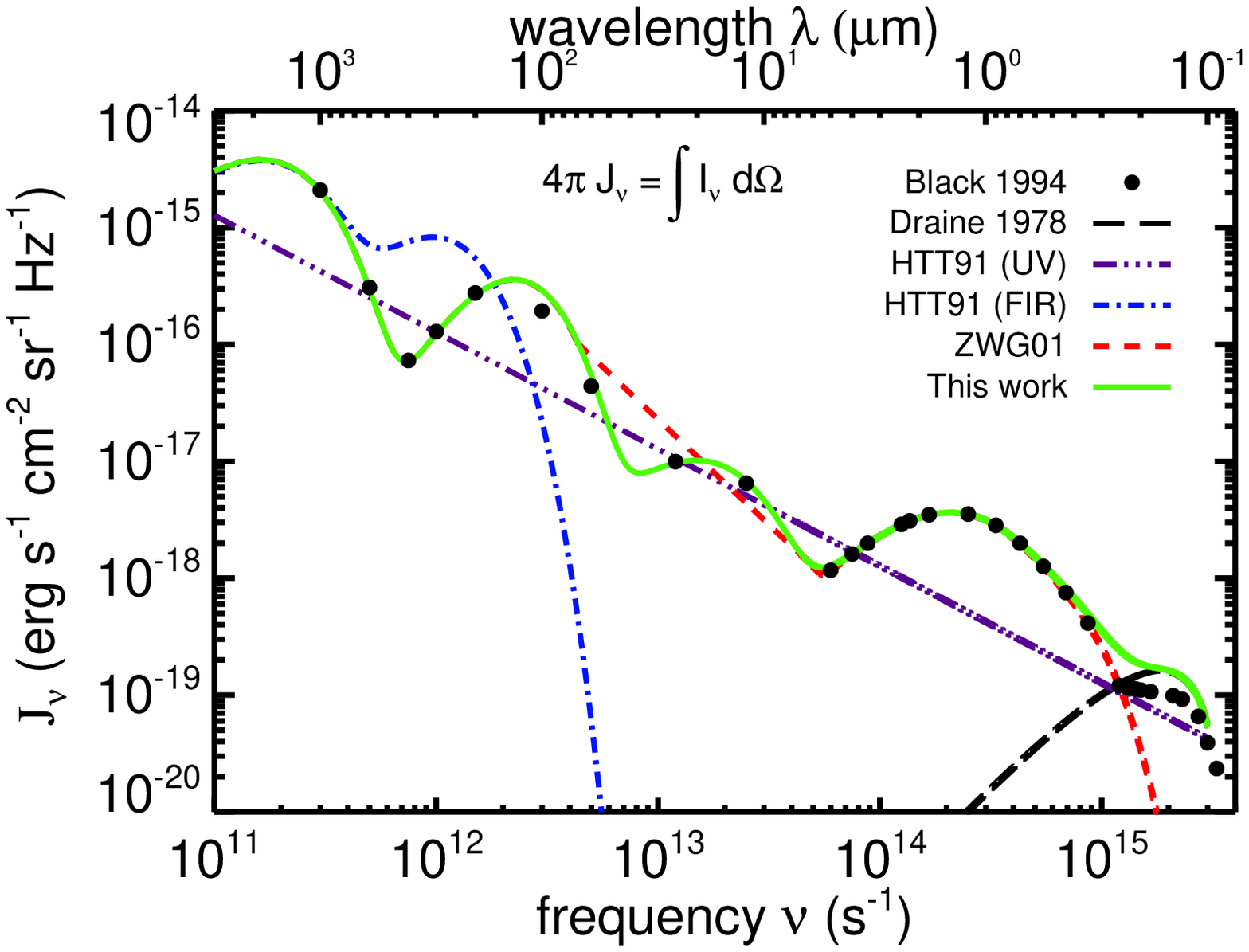} \\
\includegraphics[scale=0.51]{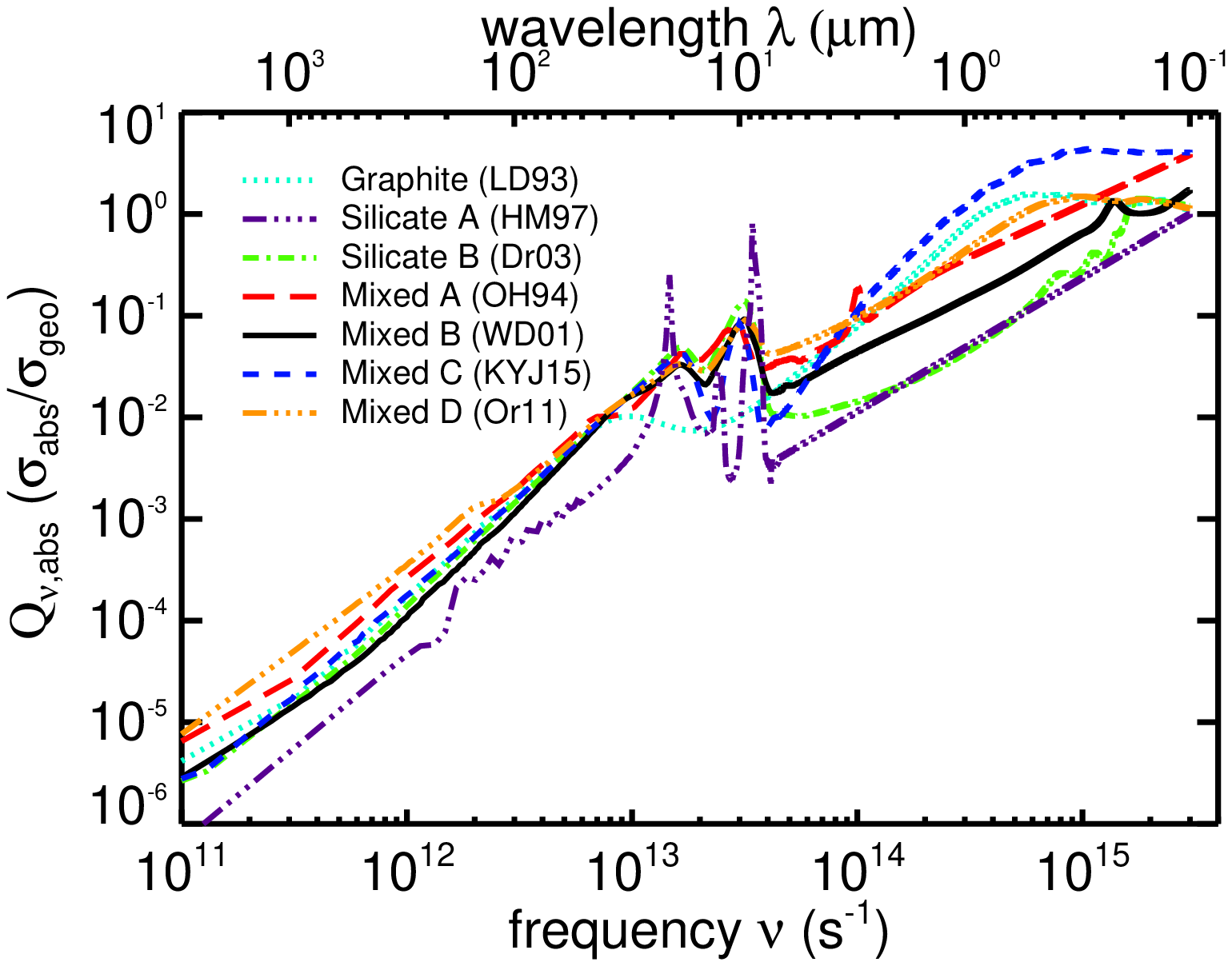}
\includegraphics[scale=0.51]{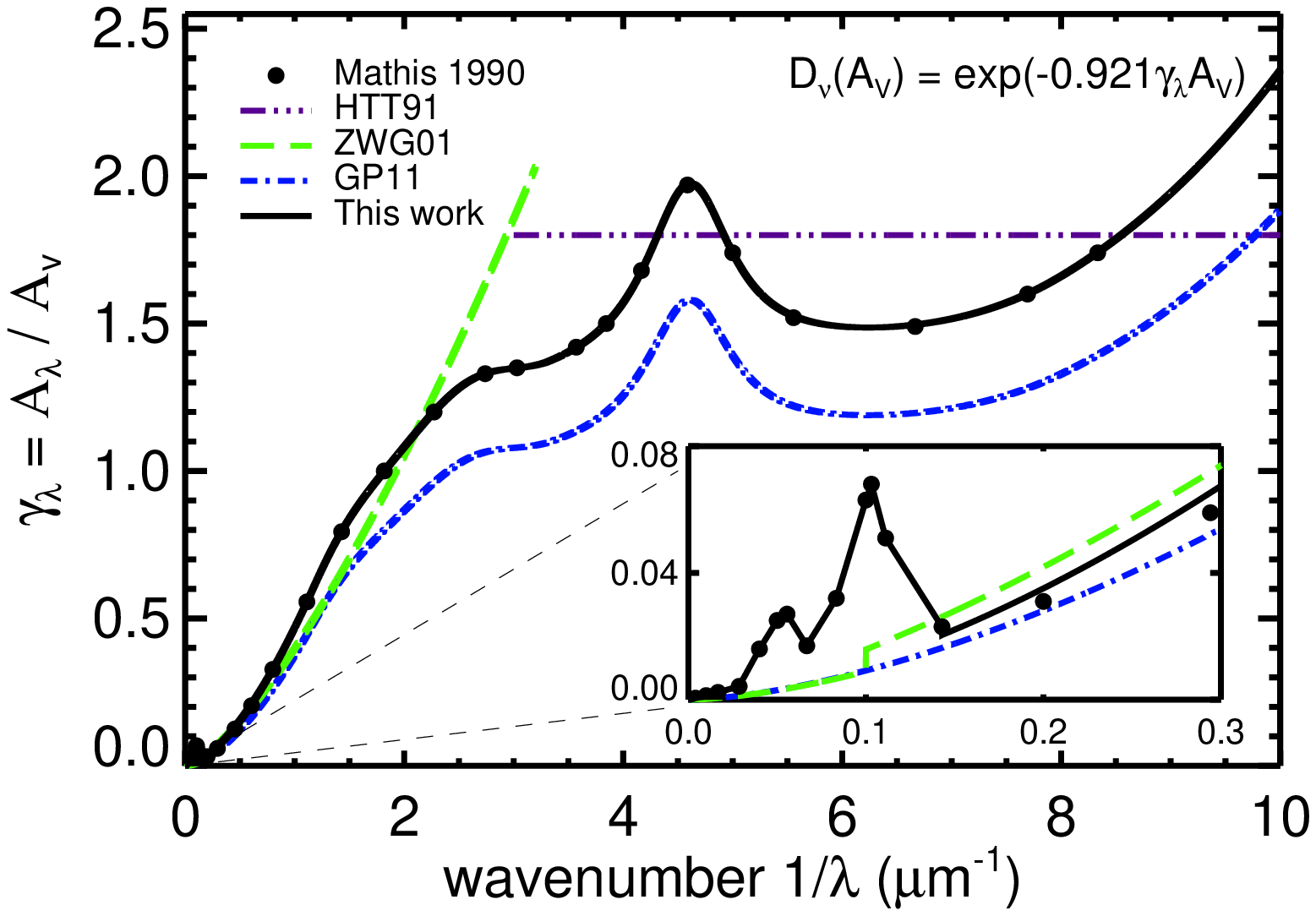}
\caption{Parameters for the dust equilibrium. Top panel, the ISRF intensity as a function of frequency. The green line represents the adopted ISRF in this work. Bottom left panel, experimental and calculated absorption efficiencies for various grain materials. Scattering is not included in these. Bottom right panel, extinction curves from various studies. The filled black circles show the observed data from \citet{1990ARA&A..28...37M}. The black solid line is a fit to the data as given by \citet{1989ApJ...345..245C}, which is adopted in this work. The sub panel zooms in at the lower wavenumbers where the adopted extinction curve below 0.15 $\mu$m$^{-1}$, given by the black solid line, is interpolated from \citet{1990ARA&A..28...37M}.}
\label{fig:eq1parameters}
\end{figure*}
Because the ISRF has contributions from various sources, it results in a continuum spectrum that goes from the FUV to the CMB. The shape and power of this spectrum is described by \cite{1983A&A...128..212M} and \cite{1994ASPC...58..355B}, though, we do note that the Galactic ISRF of most cores is anisotropic and dependent on the location within the Galaxy. In the work of \cite{2001A&A...376..650Z}, henceforth to be referred as ZWG01, the ISRF is approximated by a parametric fit to the data of the above mentioned authors. However, the UV part of the spectrum is omitted, because the aim of ZWG01 was to model the temperature at $A_{\rm V}>10$\,mag. This part of the spectrum can be covered by adopting the UV background of \cite{1978ApJS...36..595D}. This approach is also reported by \cite{2012MNRAS.421....9G} and \cite{2015MNRAS.449.2643B}. Aside from adding the UV part of the spectrum down to $\lambda=0.091$ $\mu$m (13.6 eV), we also adapt the mid-infrared (MIR) part of the spectrum to a smooth modified black body instead of the power-law with a cut-off at $100\,\mu$m of ZWG01\footnote{the ZWG01 power-law best matches the data for the range $5-70$\,$\mu$m.}. Our modified function at the MIR (around $\nu = 2\times10^{13} \rm s^{-1}$) is
\begin{equation}
B_{MIR} = \frac{2 h \nu^3} {c^2} \frac{W_i} {{\rm exp}({h \nu / k_{\rm B}T_i})-1},
\label{eq:mir}
\end{equation}
where $h$ is the Planck constant, $k_{\rm B}$ is the Boltzmann constant, $c$ is the light speed, and $W_i$ is the weighting factor. Our fitted values give us $W_i=3.4\times10^{-9}$ and $T_i=250$\,K. This approach approximates the data better than the mentioned partial power-law (see the top panel of Fig.\,\ref{fig:eq1parameters}), albeit that this part of the spectrum is actually not smooth and largely dominated by polycyclic aromatic hydrocarbon (PAH) emission, see for example, \cite{2006ApJ...648L..29P}. The full expression of the adopted ISRF is given in appendix\,\ref{app:sec:ISRF} (Eqs.\,\ref{app:eq:mir} and \ref{app:eq:uv}).

In the top panel of Fig.\,\ref{fig:eq1parameters}, we plot the ISRF as mean intensity $J_{\nu}$ ($\rm erg \,s^{-1} \,cm^{-2} \,sr^{-1} \,Hz^{-1}$) versus frequency. The black filled circles in this figure represent the original data from \cite{1994ASPC...58..355B} and \cite{1983A&A...128..212M}, which have a low UV contribution, whereas the black long-dashed line displays the \cite{1978ApJS...36..595D} UV field. The red short-dashed line shows the fit of ZWG01. We have also added in this figure the ISRF adopted by \cite{1991ApJ...377..192H}, from here on HTT91, by two separate functions: the FIR/CMB part is given in blue (dot-dashed) and the UV part in purple (triple dot-dashed). The green solid line, that is adopted in the current work, shows the combined ZWG01 and Draine \text{ISRFs}, which covers the whole frequency range, in consensus with earlier works \citep{2012MNRAS.421....9G, 2015MNRAS.449.2643B}.

\subsection{$Q_{\nu}$, the absorption efficiency}
\label{sec:Q}
\begin{table*}
\caption{Considered dust material types adopted from literature.}
\center
\begin{tabular}{lllllll}
\hline
\hline
Model & material & carbon & bulk density & literature reference & reference & data \\
& & fraction & (g\,cm$^{-3}$) & & label & link \\
\hline
\textbf{Graphite}	& carbon			& 1		& 2.26	& {\citet{1993ApJ...402..441L}} & \textbf{LD93} & 1\\
\textbf{Silicate A}	& SiO$_2$			& 0		& 3.0	& {\citet{1997A&A...327..743H}} & \textbf{HM97} & 2\\
\textbf{Silicate B}	& MgFeSiO$_4$			& 0		& 3.5	& {\citet{2003ApJ...598.1026D}} & \textbf{Dr03} & 1\\
\textbf{Mixed A}	& carbon-silicate mix	& 0.41		& 2.531	& {\citet{1994A&A...291..943O}} & \textbf{OH94} & 3\\
\textbf{Mixed B}	& carbon-silicate mix	& 0.36	& 3.2	& {\citet{2001ApJ...548..296W, 2003ARA&A..41..241D}} & \textbf{WD01} & 4\\
\textbf{Mixed C}	& carbon-silicate mix	& 0.48		& 3.05	& {\citet{2015A&A...579A..15K}} & \textbf{KYJ15} & 5\\
\textbf{Mixed D}	& carbon-silicate mix	& 0.33		& 2.65	& {\citet{2009A&A...502..845O, 2011A&A...532A..43O}} & \textbf{Or11} & 5\\
\hline
\end{tabular}
\label{tab:materials}
\tablefoot{data from: $^1$\texttt{\scriptsize http://www.astro.princeton.edu/\textasciitilde{}draine/dust/dust.diel.html}, $^2$\texttt{\scriptsize http://www.astro.uni-jena.de/Laboratory/OCDB/amsilicates.html}, \\ $^3$\texttt{\scriptsize https://hera.ph1.uni-koeln.de/\textasciitilde{}ossk/Jena/tables.html}, $^4$\texttt{\scriptsize http://www.astro.princeton.edu/\textasciitilde{}draine/dust/dustmix.html}, $^5$no online data.}
\end{table*}
Interstellar dust grains efficiently absorb photons with wavelengths smaller than their own size. Longer wavelength radiation is not entirely absorbed and there is an efficiency related to this, which is given by the frequency dependent parameter $Q_{\nu}$. Scattering is not considered in the efficiency $Q_{\nu}$ since scattered radiation will not thermally affect a dust grain. Scattering may extend the path length of a photon which increases the probability of absorption of radiation. For this, one ideally needs to keep track of all the scattered radiation at each point in a cloud. This can be done numerically. We consider scattering through the attenuation of radiation, which is discussed in Sect.\,\ref{sec:attenuation}. Scattering is very inefficient at wavelengths much larger that the size of the scattering object ($\propto \lambda^{-4}$), but may become important at wavelengths around $\lambda \lesssim 1 \mu$m.

We take $Q_{\nu}=1$ for $\lambda \ll 2\pi a$, i.e., the geometric optics approach, and $Q_{\nu} \equiv \sigma_{\rm ext}/\pi a^2$, which is less than unity, in the Rayleigh limit $\lambda \gg 2\pi a$, where $\sigma_{\rm ext}$ (cm$^2$) is the extinction cross section. The dust emissivity for cooling is an equally important aspect as dust extinction is for heating. Both are treated by the same efficiency parameter $Q_{\nu}$. It is generally acceptable to assume that the emission efficiency equals the absorption efficiency, i.e., $Q_{\nu,em} = Q_{\nu}$, since a good absorber is also a good emitter.

The absorption efficiency can be theoretically constructed, where the simpler models assume a power-law dependence, or can be experimentally measured, with material-specific absorption features. When the function follows a power-law, since $Q_{\nu}$ is on both sides of Eq.\,\ref{eq:thbalance}, only the slope of the function, i.e., the power of $\nu$, matters for the dust temperature. In reality, however, dust has material-specific absorption features. With detailed semi-analytical calculations or direct laboratory measurements it is possible to obtain the opacity coefficient $\kappa_{\nu}$ (cm$^2$ g$^{-1}$), also known as the mass absorption coefficient, with high precision. The relation between $Q_{\nu}$ and $\kappa_{\nu}$ for spherical grains is as
\begin{equation}
Q_{\nu} = \frac{4}{3} \kappa_{\nu} a \rho_d,
\label{eq:qkappa}
\end{equation}
where $\rho_d$ is the bulk mass density of dust, which is roughly around 3\,g\,cm$^{-3}$ for silicate grains, but actually depends on the dust refractory material composition. 

In the present work, we adopt a realistic set of absorption efficiencies. The opacities for the considered dust materials are gathered from the references provided in Table\,\ref{tab:materials}. The adopted absorption efficiencies are either experimentally measured or theoretically calculated from the optical properties of the refractory materials. The obtained data is in units of $Q_{\nu}$ (LD93), optical constants $n,k$ (HM97, Dr03), or $\kappa_{\nu}$ (OH94, WD01, KYJ15, Or11). The Mie theorem allows the calculation of $\kappa_{\nu}$ from $n,k$ \citep[e.g.,][]{1983asls.book.....B}, while $\kappa_{\nu}$ is converted to $Q_{\nu}$ using Eq.\,\ref{eq:qkappa}. This shows that we could just as well integrate Eq.\,\ref{eq:thbalance} over $a^3 \kappa_{\nu}$ (mass) instead of $a^2 Q_{\nu}$ (surface). Since some of the models have an underlying size distribution, the conversion with a fixed grain size makes the assumption that the canonical value of grain radius 0.1\,$\mu$m represents the mass weighted average over the grain size distribution \citep[][]{1977ApJ...217..425M, 2003pid..book.....K}. We expect that the size of the grains will not significantly change during the evolution of a diffuse cloud to a prestellar core \citep[e.g., Chacon-Tanarro et al. 2017, submitted;][]{2009MNRAS.399.1795H, 2014MNRAS.444.2303S}.

In the bottom left panel of Fig.\,\ref{fig:eq1parameters}, we show the absorption efficiencies obtained from detailed calculations and laboratory experiments for the various types of dust material. To have a matching wavelength coverage, we extrapolate the data where there are no measurements and we limit $Q_{\nu} \leq 1$, i.e., remain within the geometric optics approach (not allowing more than 100\% absorption efficiency), following \cite{1991ApJ...377..192H}.

In our selected list of materials \textsl{Graphite} is calculated for 0.1\,$\mu$m grains at 25\,K, \textsl{Silicate\,A} (quartz glass) is measured at 10\,K, \textsl{Silicate\,B} (astronomical silicate) is composed for 0.1\,$\mu$m grains at 20\,K, and \textsl{Mixed\,D} is calculated at 10\,K. For \textsl{Mixed\,A} we adopt the uncoagulated model, for \textsl{Mixed\,B} we adopt the Milky Way $R_V = 5.5$ model, where $R_V$ is discussed in the next section, and for \textsl{Mixed\,C} we adopt the uncoagulated `CMM' model. Other details can be found from the original papers.

\subsection{$D_{\nu}(A_{\rm V})$, the attenuation factor}
\label{sec:attenuation}
The attenuation of radiation in the ISM mostly arises from dust and large molecules. Depending on frequency, these particles absorb and scatter light. For radiation traveling through a medium, when calculating the optical depth $\tau_{\nu}$ considering only absorption, one has to integrate the absorption coefficient $\alpha_{\nu}$ (cm$^{-1}$) along the path $s$, that is,
\begin{equation}
\tau_{\nu} = \int_{s_0}^{s} \alpha_{\nu} ds.
\label{eq:tau}
\end{equation}
$\alpha_{\nu}$ is related to the opacity $\kappa_{\nu}$ through the relation $\alpha_{\nu}=\rho\kappa_{\nu}$. Since attenuation scales as $\exp(-\tau_{\nu})$, one needs to know the optical depth at all frequencies to find the solution for $T_{\rm d}$.

At visible wavelengths ($\lambda=5500$\AA{}) the relationship between $\tau_V$ and $A_{\rm V}$ is straightforward and the attenuation factor simplifies to $\exp(-0.921\,A_{\rm V})$. Taking this as a reference, the attenuation at different wavelengths is scaled by the wavelength-dependent attenuation coefficient $\gamma_{\lambda}$ ($\equiv A_{\lambda}/A_{\rm V}$), such that the attenuation factor $D_{\nu}(A_{\rm V})$ becomes 
\begin{equation}
D_{\nu}(A_{\rm V}) = \exp(-0.921\,\gamma_{\lambda}A_{\rm V}).
\end{equation}
The coefficient $\gamma_{\lambda}$, that is given by the extinction law, is parameterized by \cite{1989ApJ...345..245C} for the Milky Way. The extinction law accounts for both absorption and scattering, and its shape is the result of the contribution of three main components: PAHs, small carbon grains, and silicates. We use their 5-part function for the wavelength range 0.1 $\mu$m to 3.4 $\mu$m. For longer wavelengths we adopt the tabulated values of \cite{1990ARA&A..28...37M}.

The attenuation coefficient is now only a function of the optical parameter $R_V$, which is the total-to-selective extinction ratio $R_V \equiv A_{\rm V}/E(B-V)$. Two typical $R_V$'s in the ISM are for diffuse clouds with $R_V=3.1$ and for dense clouds with $R_V\sim5$. For the present work, we adopt $R_V=5$, but also discuss $R_V=3.1$ briefly. In Table \ref{tab:alav}, we show $\gamma_{\lambda}$ for a number of wavelengths.
\begin{table}
\caption{Attenuation coefficient $\gamma_{\lambda}$ ($R_V=5$).}
\center
\begin{tabular}{llllll}
\hline
\hline
$\lambda$ ($\mu$m) & $\gamma_{\lambda}$ & $\lambda$ ($\mu$m) & $\gamma_{\lambda}$ & $\lambda$ ($\mu$m) & $\gamma_{\lambda}$ \\
\hline
0.10  & 2.36  & 0.365 & 1.33  & 9.7 & 0.068 \\
0.11  & 1.97  & 0.44  & 1.20  & 10  & 0.063 \\
0.12  & 1.74  & 0.55  & 1.00  & 12  & 0.032 \\
0.13  & 1.60  & 0.7   & 0.794 & 15  & 0.017 \\
0.15  & 1.49  & 0.9   & 0.556 & 18  & 0.027 \\
0.18  & 1.52  & 1.25  & 0.327 & 20  & 0.025 \\
0.20  & 1.74  & 1.65  & 0.209 & 25  & 0.016 \\
0.218 & 1.97  & 2.2   & 0.131 & 35  & 4.2$\times10^{-3}$ \\
0.24  & 1.68  & 3.4   & 0.065 & 60  & 2.3$\times10^{-3}$ \\
0.26  & 1.50  & 5     & 0.035 & 100 & 1.3$\times10^{-3}$ \\
0.28  & 1.42  & 7     & 0.023 & 250 & 4.9$\times10^{-4}$ \\
0.33  & 1.35  & 9     & 0.051 &     & \\
\hline
\end{tabular}
\label{tab:alav}
\end{table}
The impact of $R_V$ on the extinction curve is quite large at shorter wavelengths, however, above $\lambda \simeq 0.55\,\mu$m (or $\gamma_{\lambda}<1$) the differences in $\gamma_{\lambda}$ are small \citep[$< 16$\%, for a nice overview see figure\,2 of][]{1990ARA&A..28...37M}.

In Fig.\,\ref{fig:eq1parameters} bottom right panel, we show $\gamma_{\lambda}$ as a function of wavenumber ($1/\lambda$). Here, one can notice the broad band feature at 2175\,\AA{} and the FUV rise toward shorter wavelengths. The origin of the prominent 2175\,\AA{} feature is not fully understood, but is believed that carbonaceous materials, such as PAHs \citep{2003ARA&A..41..241D, 2011ApJ...733...91X, 2012ApJ...753...82Z} or amorphous hydrocarbons \citep{2013A&A...558A..62J}, are responsible. We also overlay the extinction curves used by HTT91, ZWG01 (assuming a reference frequency of $c/5500$\,\AA), and \cite{2011ApJ...735...15G}, henceforth GP11 (assuming $R_V = 5$). In the sub panel of Fig.\,\ref{fig:eq1parameters} (bottom right), we zoom in on the lower wavenumbers to highlight the differences, which are important for embedded regions.

\section{Dust temperature: Semi-analytical solutions}
\label{sec:self}

\subsection{Bare grains}
\label{sec:mytd}
\begin{figure*}
\centering
\includegraphics[scale=0.51]{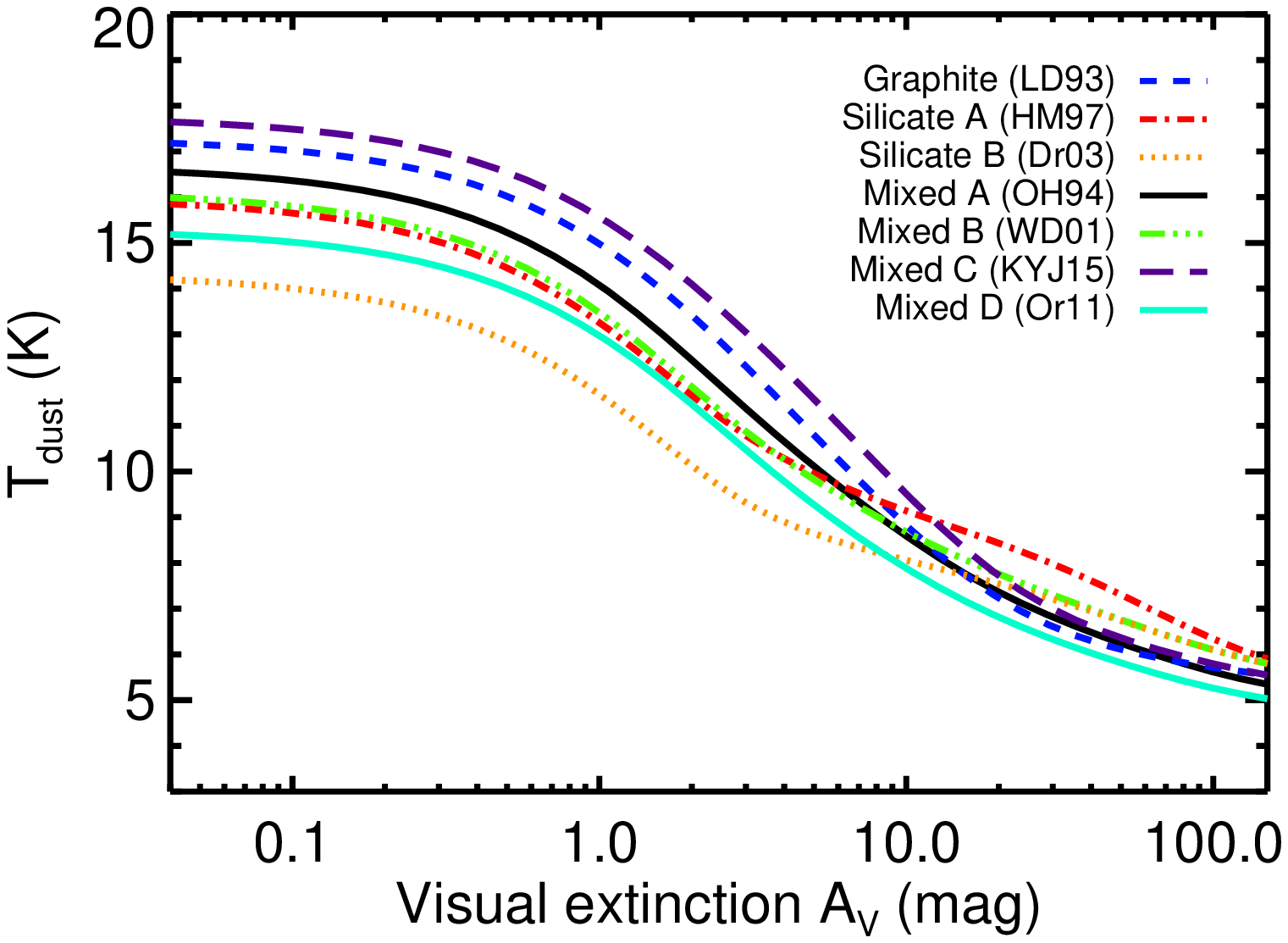}
\includegraphics[scale=0.51]{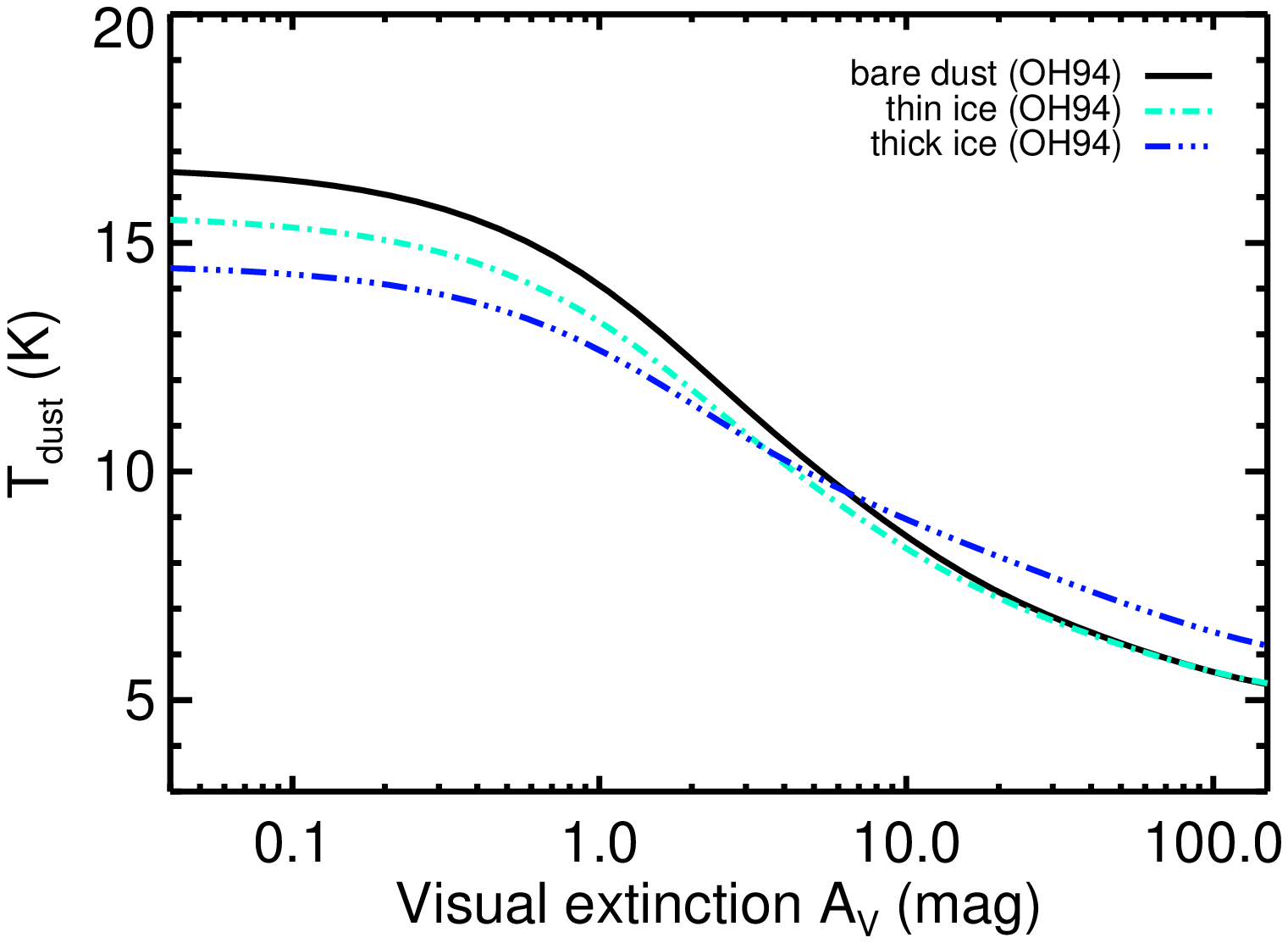}
\caption{Dust temperature solutions. The left panel displays the obtained dust temperatures for various grain materials, which have no ices. The right panel shows the impact of ice formation.}
\label{fig:td1}
\end{figure*}

We display the dust temperatures obtained by solving Eq.\,\ref{eq:thbalance} with our semi-analytical model (described in Sect.\,\ref{sec:theory}) for the materials graphite, silicate (SiO2, MgFeSiO$_4$), and carbonaceous-silicate mixtures in the left panel of Fig.\,\ref{fig:td1}. The adopted opacities for the seven different dust material types used in this work are provided in Table\,\ref{tab:materials}.

All model solutions arrive at a cloud edge ($A_{\rm V} = 0.04$\,mag) temperature that lies between $14-18$\,K. The temperature variation at high optical depths ($A_{\rm V} = 150$\,mag) is quite small for all dust types, ranging between 5 and 6\,K, where the silicate dust types are generally warmer. The silicate dust types have a smaller temperature difference between the minimum and maximum $A_{\rm V}$ and have different dependencies with extinction compared to the carbonaceous and mixed dust types. The graphitic dust temperature profile, on the other hand, is hard to differentiate from the mixed dust profiles. The larger differences that arise at lower $A_{\rm V}$ are a result of the greater variation in the absorption efficiencies at higher frequencies, see Sect.\,\ref{sec:Q}. The similar temperatures for graphite and mixed dust and the lower silicate temperature at low $A_{\rm V}$ is expected, since the opacity of carbon grains in the optical and near-infrared is higher than that of silicate grains. While for mixtures, the carbon component dominates.

\subsection{Icy grains}
We also examine the dust temperature when the grain surface is covered by (pure water) ice. Ice formation changes the opacity of dust in a way that it may reduce the dust temperature at the cloud edge, though, no ice is expected in these regions, while ice formation increases the dust temperature at high optical depths. This is because of the prominent ice features in the infrared (IR), which increase the opacity at those wavelengths. As heating effectively occurs throughout the spectrum, cooling of dust is, on the other hand, temperature dependent. Therefore, due to the ice bands in especially the near-IR (NIR) and the far-IR (FIR), the dust grain is either cooled or heated more depending on its temperature \citep[see for a review][]{2015EPJWC.10200011W}. 

OH94, KYJ15, and Or11 have also modeled grain opacities by coating dust surfaces with ices. Their method utilizes the optical properties of water ice and by applying the effective medium theory \citep[see e.g.,][]{2008A&A...489..135M, 2016A&A...586A.103W}. The OH94 modeled icy mantles range from having a few monolayers\footnote{A monolayer is when the whole dust surface is covered by one molecule thick ($\sim$\,3\,\AA) layer of ice.} of ice to $\gtrsim$\,100 monolayers. 

Using the OH94 ice models, we display in the right panel of Fig.\,\ref{fig:td1} the dust temperatures for ice covered dust grains with thick ice mantles (ice volume $\geq 4.5 ~\times$ grain volume, i.e., all water is frozen), thin ice mantles (ice volume $\geq 0.5 ~\times$  grain volume), and no ice mantles. While not drawn in this panel, the KYJ15 and the Or11 ice models also indicate the same trend, having only thin ice and bare surface models to compare. We note that both the KYJ15 and the Or11 base ice model opacities are not entirely separable from coagulation, hence we did not display them in Fig.\,\ref{fig:td1}, but we review and show them in appendix\,\ref{app:sec:ices}. We thus can conclude that ice formation has a clear and notable impact on the dust temperature, and that at $A_{\rm V}=150$\,mag, the ices make a difference of about 0.8\,K. Where thick ices result in $T_{\rm d} = 6.1$\,K at $A_{\rm V}=150$\,mag, bare grains cool to $T_{\rm d} = 5.3$\,K.

It is, however, not expected to have thick ice mantles on dust grains at low $A_{\rm V}$ ($\lesssim$\,3\,mag), since UV radiation can photodesorb ices. Moreover, the adsorption rates will be quite low at the cloud edge because of the lower densities. In dense cores, on the other hand, the expectation is that thick ice mantles will cover the cold dust surfaces where UV radiation no longer plays a significant role. In a realistic case, one should go from the black line (bare grain) in Fig.\,\ref{fig:td1}, left panel, to the blue triple-dot dashed line (thick ice) transitioning around $A_{\rm V}\sim3-6$\,mag. We emphasize that this results in a less curved, quasi linear, thermal profile. This is an important point which should be taken into account in simulations.

\section{Dust temperature: Numerical solutions}
\label{sec:radmc-3d}
In addition to the semi-analytical solutions, we calculate the dust temperatures using the same opacity models (Sect.\,\ref{sec:Q}) with the Monte Carlo radiative transfer code {\sc radmc-3d} \citep{Dullemond}. The numerical approach for solving the radiative transfer problem allows us to calculate the dust temperature in arbitrary geometries and density distributions. Despite its flexibility, it is not always suitable for large scale hydrodynamical simulations that require time-dependent solutions, because the Monte Carlo approach is computationally intensive.

There is a noteworthy fundamental difference between the semi-analytical and the numerical method. In the former case, the efficiencies adopted for the heating and cooling (through $Q_{\nu}$ in Eq.\,\ref{eq:thbalance}) differ from the opacity adopted for the attenuation of the ISRF (i.e., $D_{\nu}$ in Eq.\,\ref{eq:thbalance}). Where we choose to use the observationally obtained attenuation factor provided by \cite{1989ApJ...345..245C} and \cite{1990ARA&A..28...37M} for all our semi-analytical solutions (see Table\,\ref{tab:alav}), and thus independent of the considered opacity, \textsc{radmc-3d} uses the given dust opacities to calculate the attenuation factor self-consistently. This means that the same absorption efficiency (or opacity) is responsible for the attenuation of the ISRF. Despite its self-consistency, the latter may not be a true representation of the conditions in space, e.g, when pure materials like silicate are considered, and especially at UV wavelengths when small grains or large carbon-chain molecules, such as PAHs, are neglected.

Due to this difference, we expect moderate deviations between the methods. Furthermore, our numerical model uses a one dimensional spherical coordinate system in contrast to the plane-parallel approximation applied in the semi-analytical approach. The visual extinctions measured in one system must be converted to the other before a comparison is made \citep[see e.g.,][]{1980ApJ...236..598F, 2007A&A...467..187R}.

\subsection{The \textsc{radmc-3d} code}
\label{sec:radmc3dcode}
We utilize version 0.39 of \textsc{radmc-3d} and take advantage of the multi-threading mode of the code. We consider the same ISRF and dust opacity tables as discussed in the previous sections. The dust temperature is then calculated as a function of radius in 1D for a spherically symmetric idealized molecular cloud. The radial density distribution of the cloud, $\rho(r)$, follows a power-law profile and is given according to $\rho(r) = \rho_0 (r / R_{\rm ref})^{-2}$, where $r$, the radial position, runs from 0.1\,AU (cloud center) to 6\,pc (cloud edge). To ensure that the model probes both low and high visual extinctions, the radial coordinate grid is set logarithmically with 2000 resolution elements and finer resolution at the cloud edge. The reference radius $R_{\rm ref}$ is taken as 0.5\,pc and $\rho_0$, the density at the reference radius, is $4\times10^{-21}$\,g\,cm$^{-3}$. This is equivalent to a gas number density of $n_{\rm H} = 1\times10^{5}$\,cm$^{-3}$ with a gas-to-dust ratio of 0.01. The solutions for the dust temperature are largely independent of the parameter choices ($R_{\rm ref}$, $\rho_0$) or the profile, but gives us the necessary resolution and the dynamic range in $A_{\rm V}$ that is desired in this work.

The model cloud core is isotropically irradiated by an external radiation field as shown in the top panel of Fig.\,\ref{fig:eq1parameters} (green line). No internal heating source is considered. Besides the tracking of absorption and re-emission events of photon packages, which is inherently different from the semi-analytical calculations, \textsc{radmc-3d} is capable of taking into account (an)isotropic scattering of photons. We turn this mode off for better consistency with the semi-analytic model, but discuss and quantify the effect of scattering in Sect.\,\ref{sec:scattering}. To reduce the intrinsic statistical noise of the Monte Carlo radiative transfer method, we set the number of propagated photon packages to $10^7$, which is a relatively high number for an effectively 1D model. The estimated error will be on the order of 1/$\sqrt{N_{\rm photons}}$, where the photons will be spread out among the resolution elements causing the highest error to be at the core.

The \textsc{radmc-3d} code takes the opacity $\kappa_{\nu}$ instead of the absorption efficiency $Q_{\nu}$ as input parameter. We convert between the quantities according to Eq.\,\ref{eq:qkappa} and fix the grain radius to 0.1\,$\mu$m as we do for the semi-analytical calculations. The considered opacity model types are listed in Sect.\,\ref{sec:Q}.

The code gives the radial distribution of dust temperatures for the opacity models as result. The radial position is converted to visual extinction by calculating the visual optical depth at each location from the cloud edge. The visual extinction at each radial position of the model grid is defined according to
\begin{equation}
A_{\rm V} = 1.086 \, \tau_{5500 \rm\mathring{A}},
\end{equation}
where $\tau_{5500 \rm\mathring{A}}$ is the optical depth at $\lambda = 5500$\,\AA{} and is given by Eq.\,\ref{eq:tau}. In our model, $\kappa_{5500 \rm\mathring{A}}$ is independent of the position in the cloud and can, therefore, be brought out of the integral. The integral then simplifies to a summation of the dust column density, whereas the optical depth is given by the product of the column density and the opacity. In order to have a one-to-one comparison with the semi-analytical models, we need to rescale the $A_{\rm V}$ to account for the geometry, i.e., spherical geometry to semi-infinite slab. The rescaling factor is given by \cite{2007A&A...467..187R}:
\begin{equation}
A_{\rm V, \rm eff} = -\ln \left( \int_0^1 \frac{ \exp(-\mu\tau_{\nu}) }{\mu^2} {\rm d}\mu \right) \frac{A_{\rm V}}{\tau_{\rm UV}},
\end{equation}
where $\mu=\cos\theta$ is the cosine of the radiation direction and $\tau_{\rm UV}$ is the optical depth at UV, evaluated at $0.3\,\mu$m in the present work.

\subsection{Numerically obtained dust temperatures}
\label{sec:radmc3dresults}
\begin{figure}
\includegraphics[scale=0.51]{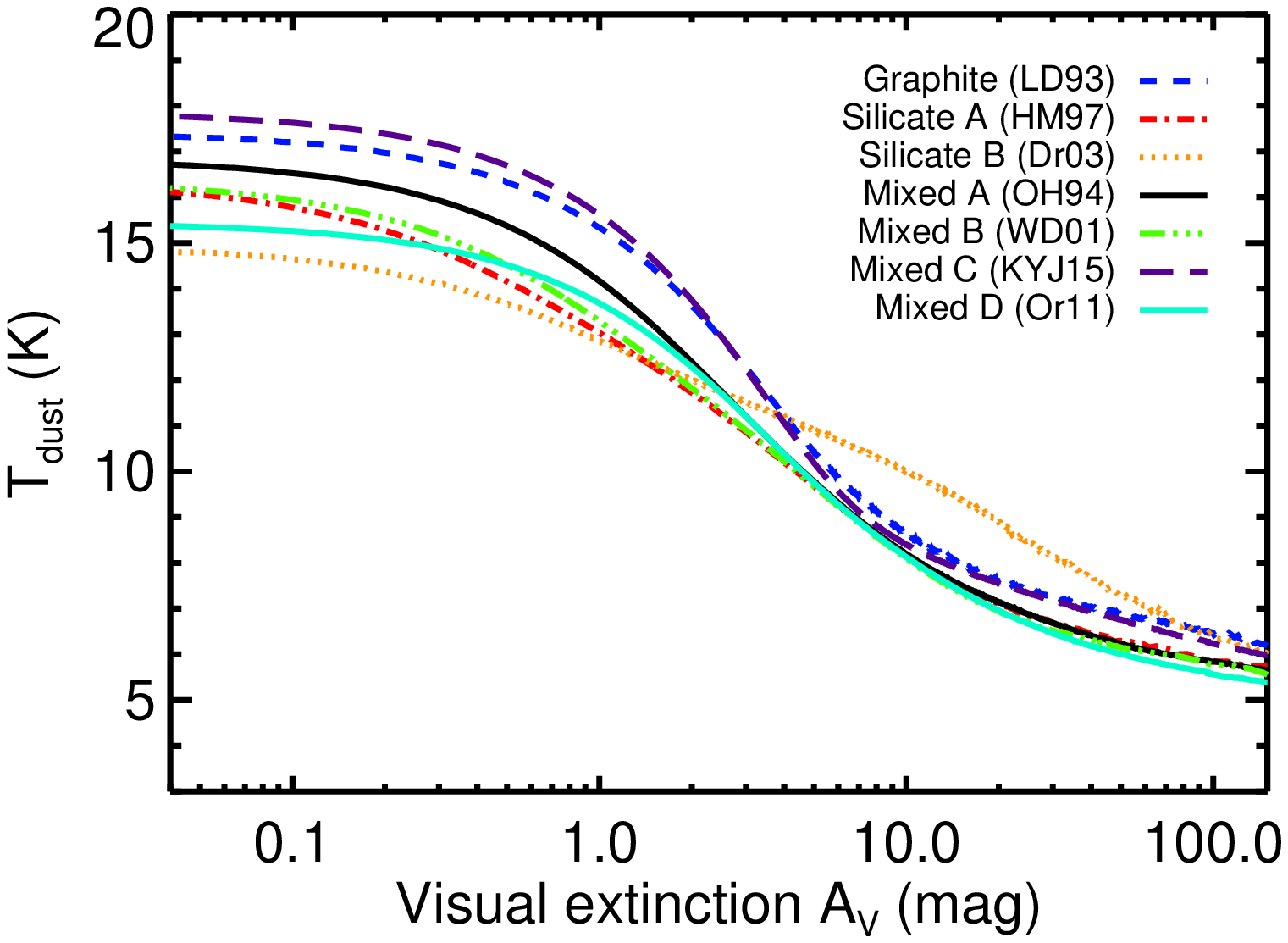}
\caption{Dust temperature solutions obtained with the \textsc{radmc-3d} code for various grain materials as given in Table\,\ref{tab:materials}.}
\label{fig:radmc1}
\includegraphics[scale=0.51]{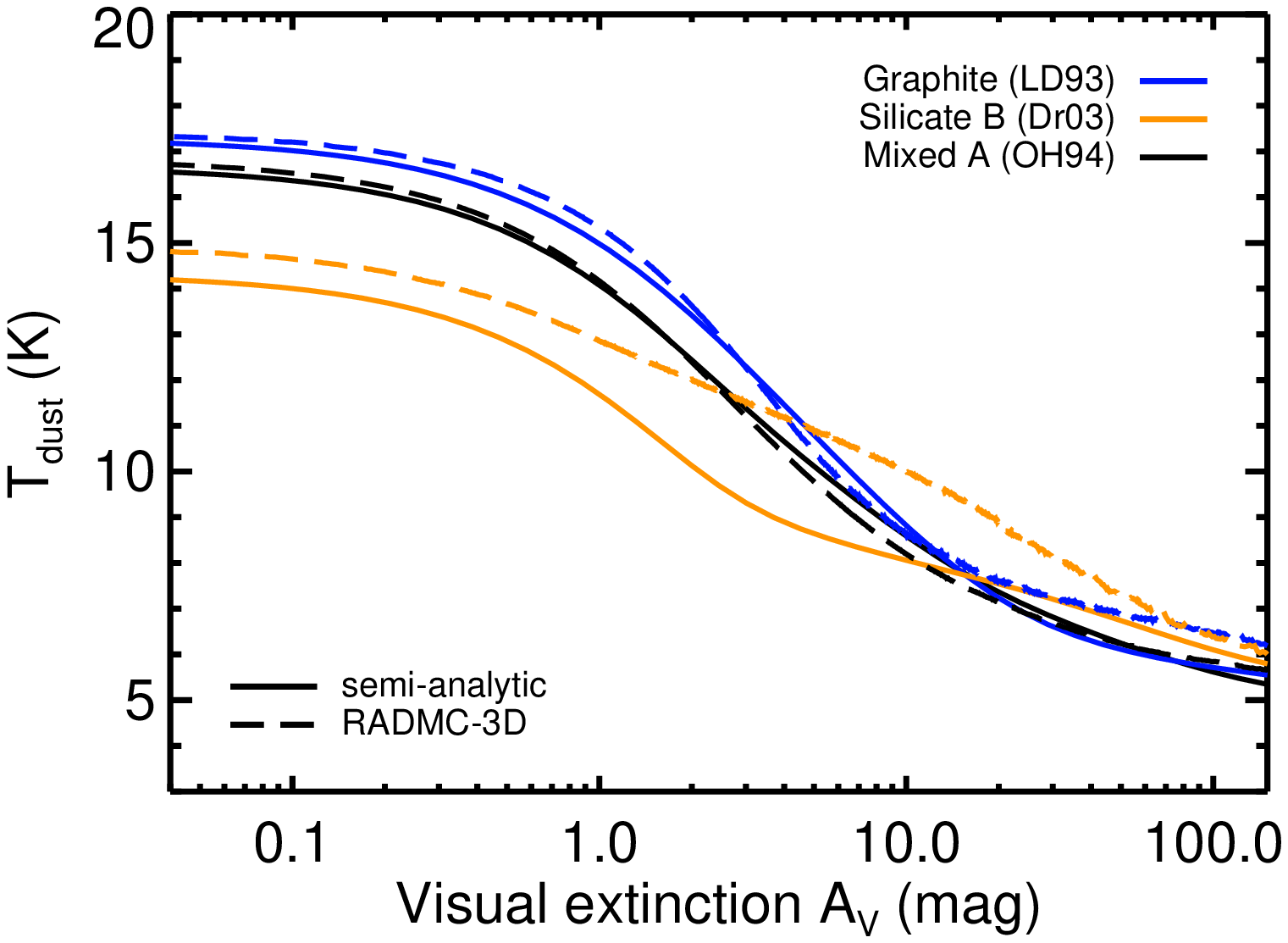}
\caption{Comparisons between the solutions obtained with the numerical (dashed lines) and the semi-analytical models (solid lines) for three different types of grain. The differences in temperature originates mainly from the underlying attenuation coefficient.}
\label{fig:radmcvsanalytical}
\end{figure}

We show in Fig.\,\ref{fig:radmc1} the dust temperature solutions obtained with the \textsc{radmc-3d} code for the dust materials graphite, silicate, and carbonaceous-silicate mixtures.
We find similar trends and draw the same conclusions for the different dust materials as we did with our semi-analytical solutions. This means that the silicate dust types have, in general, a lower dust temperature at $A_{\rm V}<2$\,mag and a higher one at $A_{\rm V}\geq150$\,mag as compared to the carbonaceous materials, which results in a smaller difference in temperature between the two boundaries. The temperature variation between all models at the edge ($A_{\rm V}=0.04$\,mag) ranges from 14.8 to 18\,K, similar to the semi-analytical solutions, and when drawn further (to $A_{\rm V} \sim 0$) they completely match the semi-analytical solutions. The agreement also holds at high depths, that is, $A_{\rm V} \geq 150$\,mag.

The main differences as compared to the semi-analytical models come from the silicate grains. The temperature profiles of the silicate grains are shaped differently from the semi-analytical solutions. The silicate dust temperatures from \textsc{radmc-3d} can differ by up to 2.4\,K, though mostly is less than 1\,K. We attribute the differences obtained with \textsc{radmc-3d} to the attenuation of radiation, which is self-consistent instead of using the observed extinction curve. Therefore, for pure silicates, this misses the important carbon opacity contribution at short waves. We compare the differences in extinction curves in appendix\,\ref{app:sec:extinction}, which clarifies the found discrepancies. As expected, the results from the two different methods do not match for pure silicate dust opacities, whereas they do match well for the mixed dust materials, especially \textsl{Mixed\,A} (OH94). The \textsc{radmc-3d} dust temperature solution for the \textsl{Mixed\,A} dust is practically identical to the semi-analytical solution. For three out of the seven models, i.e., \textsl{Graphite}, \textsl{Silicate\,B} (worst match), and \textsl{Mixed\,A} (best match), we show and highlight the differences in Fig.\,\ref{fig:radmcvsanalytical}.

\subsection{Scattering}
\label{sec:scattering}
The numerical approach allows us to change the geometry and include the scattering of interstellar photons on dust grains. We test and discuss the results at higher spatial dimension geometries in appendix\,\ref{app:sec:geometry}. While the choice of geometry does not greatly impact the results, the expectation is that scattering may play a more appreciable role, especially for the attenuation of energetic photons (at short wavelengths). Here we discuss the impact of scattering.

\begin{figure}
\includegraphics[scale=0.36]{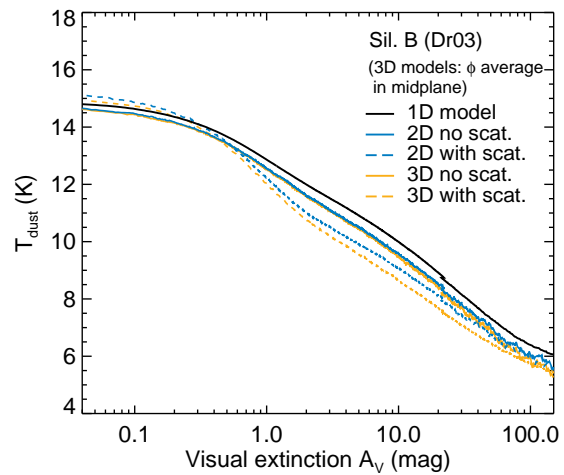}
\caption{The dust temperatures with scattering. The dashed lines show the \textsc{radmc-3d} solutions (2D in blue, 3D in yellow) when scattering is treated. The midplane ($\theta=0$) values are adopted and for 3D, the median of $\phi$ is taken.}
\label{fig:scattering}
\end{figure}

Because we do not have the scattering opacities for all the dust models, the model \textsl{Silicate\,B} is alone computed and displayed as an example. From Fig.\,\ref{fig:scattering} we can see that the impact of scattering is such that the dust temperature rises at the cloud edge, while the temperature decreases at higher $A_{\rm V}$. 
The differences are larger for 3D than for 2D. This is easily explained by the fact that due to the higher dimensions, the path length of a photon increases (the photon can travel in more dimensions) and, therefore, gets more extincted and absorbed at low $A_{\rm V}$. This causes the lower $A_{\rm V}$ temperature to be higher and the higher $A_{\rm V}$ temperature to be lower. By not considering scattering, the underestimation from the 1D model is about 0.3\,K ($\sim$2\%) at low $A_{\rm V}$, whereas the overestimation, peaking at $A_{\rm V} = 5$\,mag, is 1.5\,K ($\sim$14\%). At maximum $A_{\rm V}$ (150\,mag), the difference between 1D and 3D is about 0.8\,K ($\sim$13\%). We point out, once again, that in the semi-analytical calculations, scattering is considered through the extinction curve (see Sect.\,\ref{sec:attenuation}).

\section{New parametric expression for $\rm T_{d}$}
\label{sec:expression}
\begin{figure}
\includegraphics[scale=0.51]{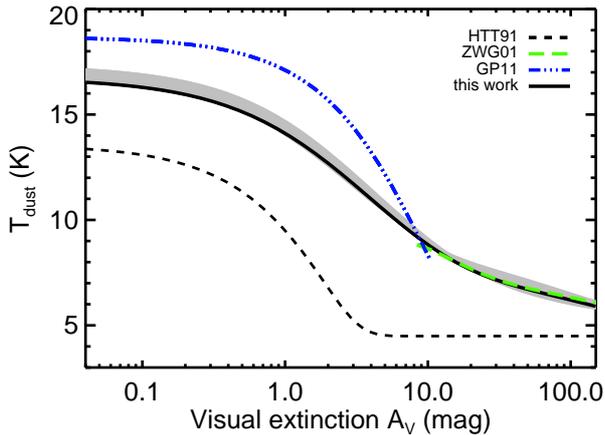}
\caption{The $T_{\rm d}^{\rm Hoc}$ parametric expression as a function of $A_{\rm V}$. We present our best fit to the theoretical calculations in this work and compare it to other parametric expressions found in the literature. The grey band illustrates the variation we get from considering thin or thick ices or from using three models to fit our expression instead of four.}
\label{fig:bestfit}
\end{figure}

From the theoretical models that best approximate the observational results (these are the models that include the mixed dust material compositions, i.e., OH94, WD01, KYJ15, Or11) we create a simple and useful expression for the dust temperature. This is achieved by finding a function to match the model results. We find that the correlation is best reproduced by fitting a hyperbolic curve through the semi-analytical solutions. Our expression is formulated as
\begin{eqnarray}
T_{\rm d}^{\rm Hoc} = \left[ 11 + 5.7\times \tanh\bigl( 0.61 - \log_{10}(A_{\rm V})\bigr) \right] \, \chi_{\rm uv}^{1/5.9},
\label{eq:expression}
\end{eqnarray}
where the expression is scalable with $\chi_{\rm uv}$, the Draine UV field strength \citep{1978ApJS...36..595D}. This matches our $F_{\rm ISRF}$, the interstellar radiation field flux, where the mean flux in the radiation field with energy between $6-13.6$\,eV is $G_0$ times $1.6\times10^{-3}$\,erg\,cm$^{-2}$\,s$^{-1}$. For the Draine field, $G_0 = 1.7$. In Fig.\,\ref{fig:bestfit}, we compare our expression with the expressions from past studies \citep[i.e.,][]{1991ApJ...377..192H, 2001A&A...376..650Z, 2011ApJ...735...15G}, which we describe in detail in appendix\,\ref{app:sec:literature}. In greyscales we highlight the variation that we get from our expression if we consider only thick (instead of thin) ices at $A_{\rm V} \geq 20$\,mag, or when we exclude the ice model, i.e., \textsl{Mixed\,D} (the turquoise line as given in the left panel of Fig.\,\ref{fig:td1}).

It is interesting to note that we find a $\chi_{\rm uv}$ (or $F_{\rm ISRF}$) with a power of 1/5.9 to best match the solutions, in close agreement with the analytical prediction of 1/6 by \citet[][see his equation 24.19]{2011piim.book.....D}. The given formula is tested and validated for the range $A_{\rm V} = 0.01 - 400$\,mag and $\chi_{\rm uv} = 0.1 - 10^5$\,erg\,cm$^{-2}$\,s$^{-1}$. The latter is shown in appendix\,\ref{app:sec:uv}. To account for ices, we fitted our line through bare dust results in the range $0 \leq A_{\rm V} \leq 6$\,mag and icy dust for $A_{\rm V} > 6$\,mag (taking thick ices from OH94). This expression is constructed for an $R_V$ of 5, but we find only small differences when compared against $R_V = 3.1$. The fitting function for $R_V = 3.1$ is given in appendix\,\ref{app:sec:fits}, where the expression is also provided for other parameter choices.

\section{Discussion: Observed versus Theoretical T$_{\rm d}$}
\label{sec:combine}
\begin{figure*}
\centering
\includegraphics[scale=0.66]{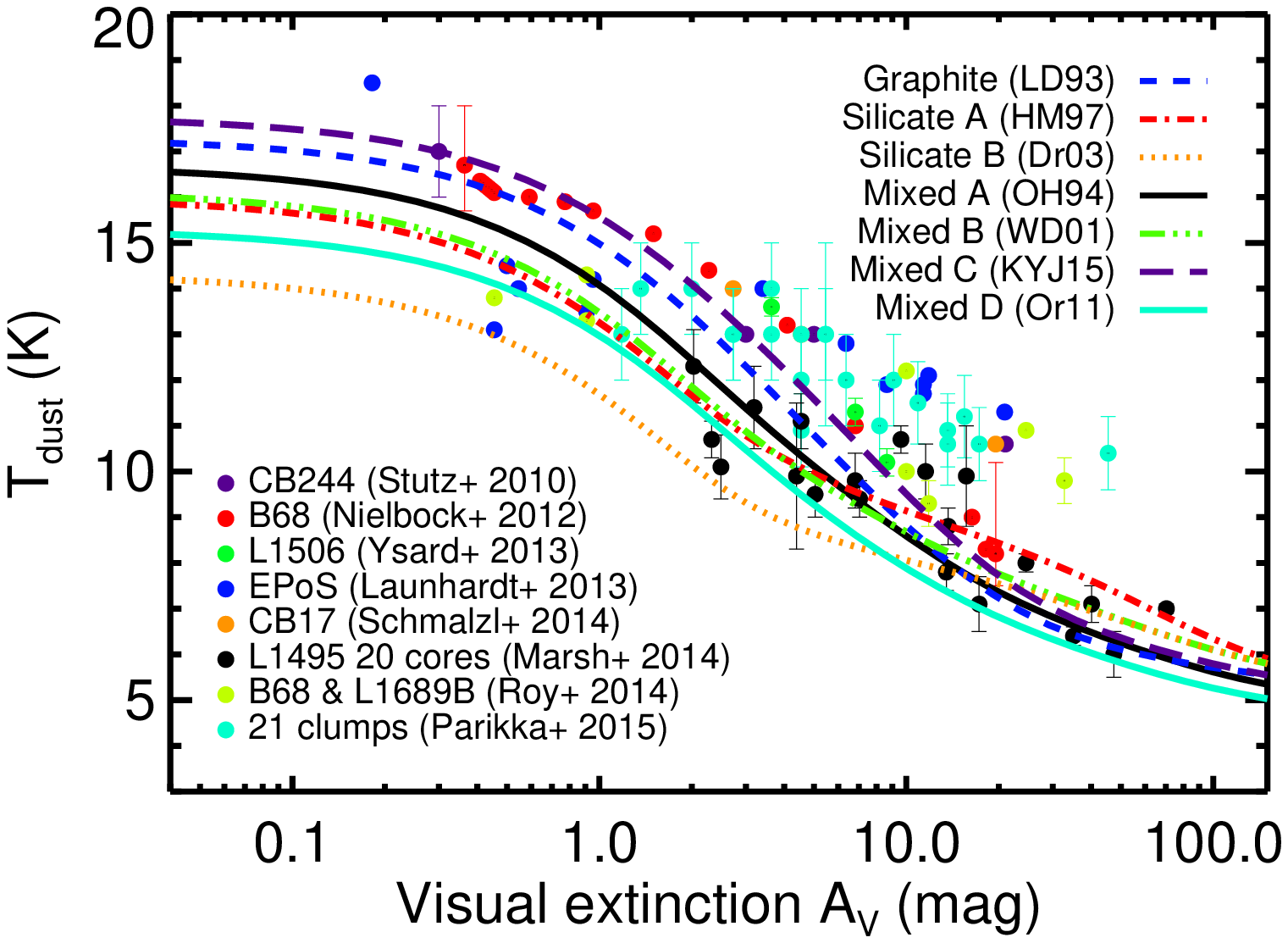} \\
\includegraphics[scale=0.51]{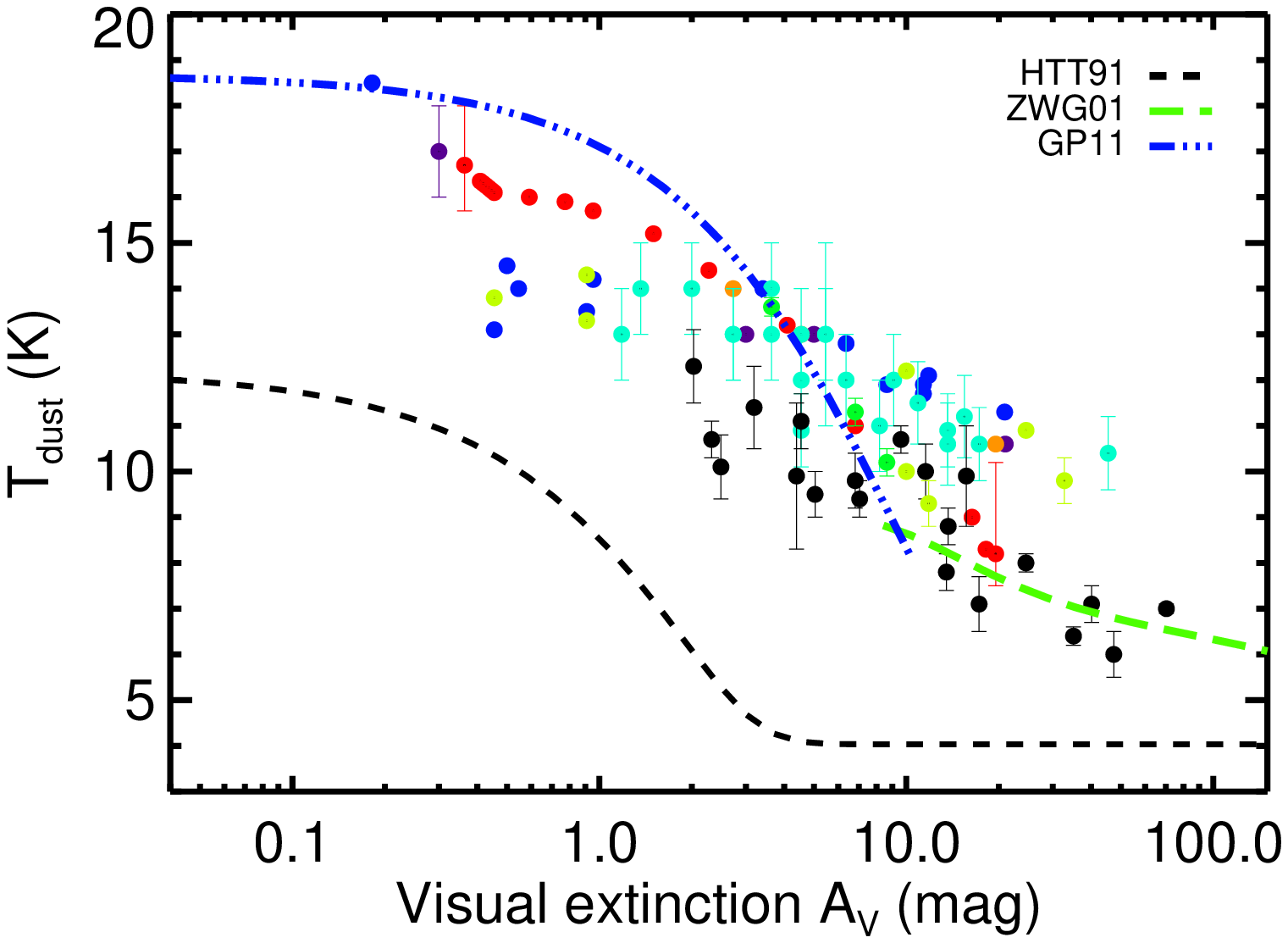}
\includegraphics[scale=0.51]{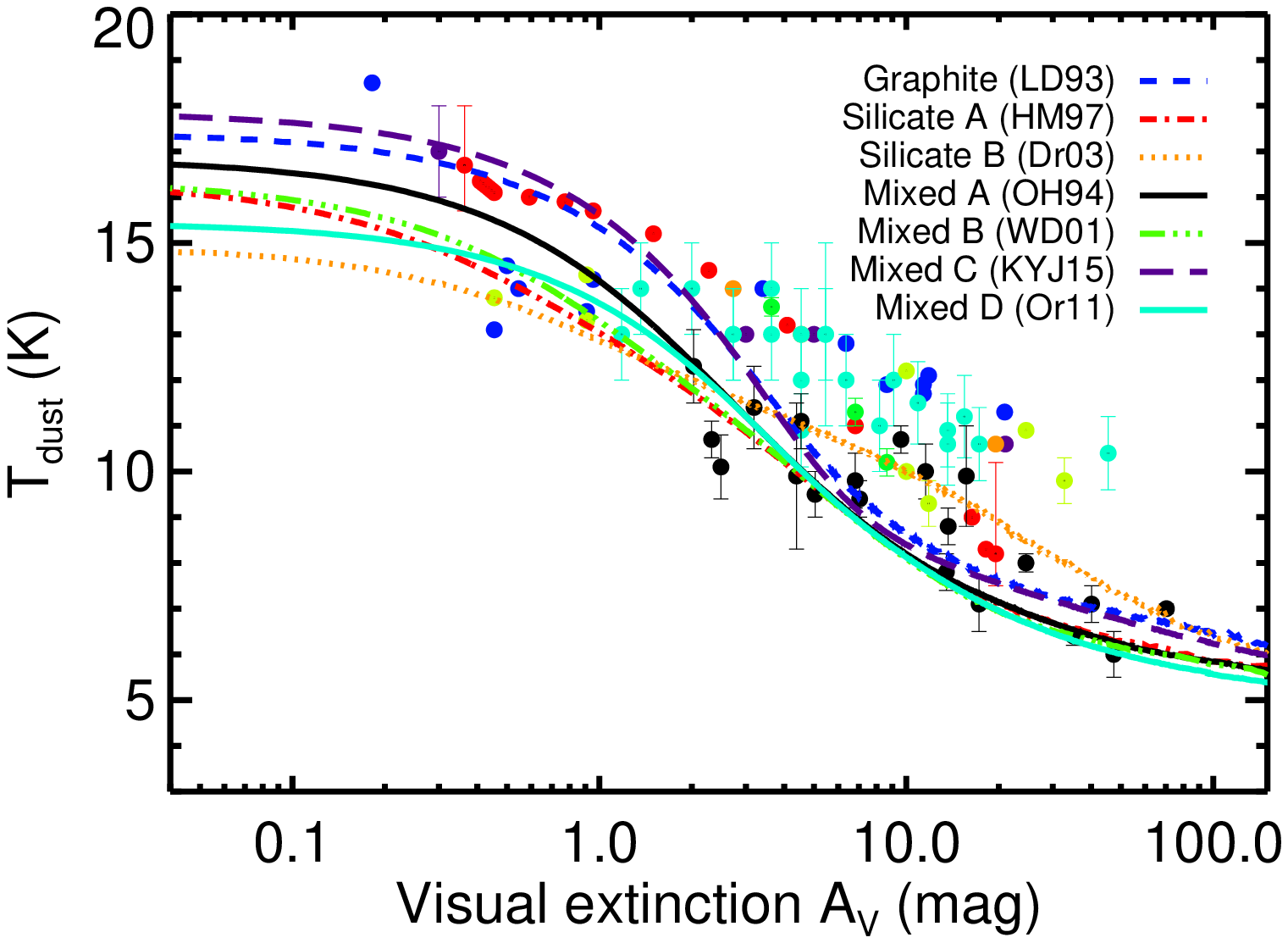}
\caption{Dust temperature solutions compared against observations. Top panel displays the obtained dust temperatures from semi-analytical solutions (Sect.\,\ref{sec:self}) for various grain materials, without ices. The bottom left panel shows the observed dust temperatures against three parametric expressions found in the literature (appendix\,\ref{app:sec:literature}). The bottom right panel compares the observed dust temperatures against the \textsc{radmc-3d} solutions (Sect.\,\ref{sec:radmc-3d}).}
\label{fig:td4}
\end{figure*}

We compare the observationally derived dust temperatures against the theoretical solutions and show them in Fig.\,\ref{fig:td4}.
Here, we discuss our main findings with respect to observations.

\begin{figure}
\includegraphics[scale=0.51]{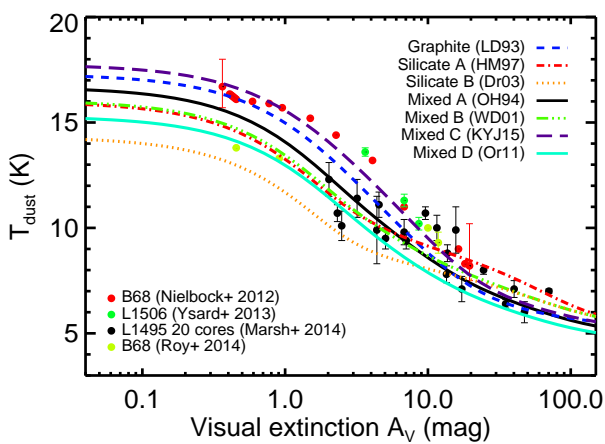}
\caption{Semi-analytical solutions compared against the LOS corrected observed dust temperatures. Only sources consistent with $G_0 \sim 1$ are considered. The match improves markedly.}
\label{fig:tdnoLOS}
\end{figure}

\textbf{Versus semi-analytic.} Our solutions provide a good match to the observational data as can be seen from the top panel of Fig.\,\ref{fig:td4}. Both the curvature as well as the values are captured well. Nevertheless, there appears to be a spread in the observational dust temperatures, particularly around $A_{\rm V} \sim 10$\,mag. This may hint toward a spread in the underlying dust material (affecting opacity) or grain size distribution among sources, however, the uncertainties in the physical conditions and, especially, the ISRF may also just be the cause. Furthermore, when ices start to cover the surface of the dust, the composition of the refractory components tend to become less relevant. Indeed, when only selecting the LOS corrected observations and excluding the higher $G_0$ environments, i.e., L1689B and CB17 \citep[][]{2014A&A...562A.138R, 2014A&A...569A...7S}, the match improves greatly, as shown in Fig.\,\ref{fig:tdnoLOS}.
This indicates that the uncertainties arising from $Q_{\nu}$ at $A_{\rm V} > 10$\,mag, i.e., when not considering ices, does not influence the dust temperature significantly.

We find that all models match the data reasonably, but the best match, with the lowest $\chi^2$, is \textsl{Mixed\,C} (KYJ15), the amorphous carbon-silicate mix with a carbon volume fraction of 0.48. This model has the highest carbon fraction among the considered mixed dust types. We also find that if we take ices into account in our models above an $A_{\rm V}$ of 6\,mag, the correlation to the observations improves.

\textbf{Versus literature.} The parametric expressions from the selected literature studies do not agree well with the temperatures acquired observationally, with the exception of ZWG01 at $A_{\rm V} \gtrsim 10$\,mag (Fig.\,\ref{fig:td4} bottom left). Where the considered literature expressions were derived for a fixed parameter space, within the bounds of their respective studies, the expression attained from this work has a broad range of validity in $A_{\rm V}$ and $G_0$. We find that the greatest difference among the literature expressions arises from the adopted absorption efficiencies, i.e., $Q_{\nu}$, which in all cases are power-law functions of frequency in the mentioned studies. Our calculations are an improvement in this domain, since we are not committed to a strict power-law, but rather adopt the measured and the intricately calculated opacities.

\textbf{Versus \textsc{radmc-3d}.} The temperatures obtained with the radiative transfer code are slightly lower, yet similar to the semi-analytical ones. Compared to observations, the dust temperatures obtained with \textsc{radmc-3d} seem to agree well especially at low $A_{\rm V}$, but are generally below the observed $T_{\rm d}$'s at $A_{\rm V} > 2$\,mag, that is, except for \textsl{Silicate\,B} (Dr03). This indicates that using the observed extinction curve for the attenuation of radiation may indeed provide a more accurate picture at $A_{\rm V} > 2$\,mag than the self-consistent extinction from the adopted opacities as \textsc{radmc-3d} does. Moreover, along the LOS a mixture of dust types and sizes will eventually be responsible for the attenuation. Ideally, one should modify the opacity input for radiative transfer calculations to take the components of the extinction curve (PAHs, carbon grains and silicates) into account. It is, on the other hand, also true that the dust temperatures from observations have the difficulties of the LOS. Even with the corrections, these may cause an overestimation of the dust temperature, especially in embedded regions \citep[cf.][]{2006ApJ...640L..47S, 2008ApJ...684.1228S, 2014A&A...562A.138R}. Thus, a lower dust temperature as compared to the observations at higher $A_{\rm V}$ is expected. 


\textbf{Versus active star-forming regions.} Our expression for the dust temperature, i.e., $T_{\rm d}^{\rm Hoc}$, scaled by the ISRF (see appendix Fig.\,\ref{app:fig:multiG0}), is in agreement with the observed dust temperature of $\rho$\,Oph\,A. For an environment with a $10^3\,G_0$ and at $A_{\rm V}=20$\,mag, our expression gives a dust temperature of 22.3\,K, similar to the results obtained from line emission of H$_2$O$_2$ \citep[$22\pm3$\,K,][]{2011A&A...531L...8B} and not far from HO$_2$ measurements \citep[$16\pm3$\,K,][]{2012A&A...541L..11P}. While for an $A_{\rm V}=100$\,mag, we obtain 17.6\,K from our expression, similar to the results retrieved from D$_2$CO measurements of $\rho$\,Oph\,A \citep[17.4\,K,][]{2011A&A...527A..39B}.

\section{Conclusions}
\label{sec:conclusion}
The motivation of this study was to find and provide a parametric expression for the dust temperature to use in numerical simulations and chemical models. To this end, we calculated the dust temperature from basic principles for dust in thermal equilibrium by considering in detail the ISRF, the attenuation of radiation, and the dust opacities. We did this for various grain material compositions, i.e., for graphite, silicates SiO$_2$ and MgFeSiO$_4$, and carbonaceous silicate mixtures. We compared our calculations against solutions obtained with the Monte Carlo radiative transfer code \textsc{radmc-3d} and against recent observational results from \textit{Herschel} that we collected from the literature.

We find that our semi-analytical solutions as well as our numerical solutions match the range of observed dust temperatures well at low and at high optical depths and also captures the overall extinction dependence (barring uncertainties in the ISRF). Mixed carbonaceous silicate dust material compositions match the observed temperatures of starless regions better than pure materials and give a narrow range temperature solution between 15.2 and 17.7\,K at the edge, that is $A_{\rm V} = 0.04$\,mag. This conforms to $5-6$\,K around an $A_{\rm V} = 150$\,mag. However, the dust surface should be covered by ices at high $A_{\rm V}$ making the composition of the refractory components less relevant in this regime.

Considering the impact of ices, we find that ice formation changes the opacity of dust significantly enough to reduce the net cooling at $A_{\rm V} > 10$\,mag. This allows the dust to be slightly warmer ($\sim$15\% for thick ice and around 8\% for thin ice) in highly embedded regions, which may be crucial in avoiding the freeze-out of H$_2$ molecules in models (typically below $T_{\rm d} \simeq 7$\,K). 
Ice formation helps to raise the dust temperature at high optical depths ($A_{\rm V} \gtrsim 20$\,mag) by about 0.5 to 1\,K, depending on ice thickness. 
The ices also aid in flattening the temperature profile, which helps in explaining the near (semi-log) linear profile inferred from the observed dust temperatures. We find that with ices (at $A_{\rm V} \gtrsim 6$\,mag) our models give a better match to the observed dust temperatures. 

From our best matching lines, we provide an analytical expression as a function of $A_{\rm V}$, given by Eq.\,\ref{eq:expression} (detailed further in appendix\,\ref{app:sec:fits}), which can be scaled by the ISRF.

\begin{acknowledgements}
We thank M. K\"ohler, N. Ysard, and C. Ormel for providing their opacity data tables. We thank A. Ivlev for his contribution on cosmic rays and M. R\"{o}llig for sharing his geometry conversion code. SH especially thanks J. Steinacker and W.F. Thi for their great interest and constructive comments about this work. PC, MS, and SC acknowledge the financial support of the European Research Council (ERC; project PALs 320620). SC is also supported by the Netherlands Organization for Scientific Research (NWO; VIDI project 639.042.017).
\end{acknowledgements}

\bibliography{Hocuk_Tdust_2017.bib}

\appendix

\section{Scaling with cosmic rays}
\label{app:sec:crdr}
\begin{figure}
\includegraphics[scale=0.51]{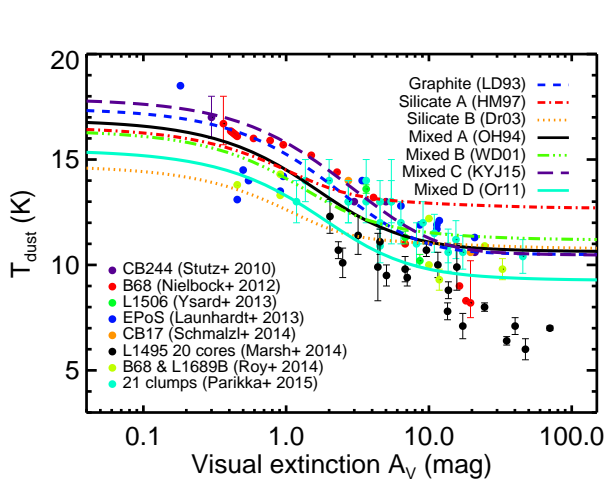}
\caption{The dust temperatures from semi-analytical solutions for a 1000 times higher cosmic ray rate.}
\label{app:fig:crdr}
\end{figure}

As a test to the impact of higher CR rates, for example in CR-dominated regions \citep[e.g.,][]{2011MNRAS.414.1705P, 2011MNRAS.414.1583B}, the calculations are also performed with elevated CR rates. In Fig.\,\ref{app:fig:crdr} the dust temperature solution for a 1000 times higher CR rate, i.e., $\zeta_{H_2} = 5\times10^{-14}$\,s$^{-1}$ is presented.
As noted before, only the heating by UV photons created from CRs are considered, not the direct CR impact or other CR processes.

With an increased CR rate, the resulting dust temperatures above an $A_{\rm V} \approx 5$\,mag are much higher, while at lower extinctions this is not the case. CRs essentially set a floor temperature for the dust grains, below which they cannot cool, though, this could be overcome by the magnetic mirroring process \citep{2013ASSP...34...61P}. It is interesting to note that the higher grain temperatures, that is seen from the observations around an $A_{\rm V} \sim 10$\,mag, can be reproduced by assuming a 1000 times higher CR rate.

\section{The ISRF}
\label{app:sec:ISRF}
The full expression for the ISRF that is used in the calculations of this work is reported here. The five modified black bodies from the work by \cite{2001A&A...376..650Z} is adopted, except for the MIR range which is changed as was presented in Eq.\,\ref{eq:mir} into a sixth modified black body. To this function the UV part of the spectrum is added. The part of the spectrum without UV is given by the six modified black bodies as
\begin{equation}
J_{\nu}^{\rm no UV} = \frac{2 h \nu^3} {c^2} \sum_i \frac{W_i} {{\rm exp}({h \nu / k_{\rm B}T_i})-1},
\label{app:eq:mir}
\end{equation}
where the values for $W_i$ and $T_i$ are given in Table\,\ref{tab:app-tab1}.
\begin{table}
\caption{Parameters for the ISRF}
\begin{tabular}{ccc}
\hline
\hline
$\lambda$ ($\mu$m) & $W_i$  & $T_i$ (K) \\
\hline
0.4	& 1$\times10^{-14}$  & 7500  \\
0.75	& 1$\times10^{-13}$  & 4000  \\
1	& 4$\times10^{-13}$  & 3000  \\
10	& 3.4$\times10^{-9}$ & 250   \\
140	& 2$\times10^{-4} $  & 23.3  \\
1060	& 1		     & 2.728 \\
\hline
\end{tabular}
\label{tab:app-tab1}
\end{table}
The included values for the MIR match the 10 micron emission coming from hot dust, but is highly smoothed out. The UV part of the spectrum is adopted from \cite{1978ApJS...36..595D}, rewritten in the current form to match the units, that is given as
\begin{equation}
J_{\nu}^{\rm UV} = 4280\big(h\nu\big)^2 - 3.47\times10^{14}\big(h\nu\big)^3 + 6.96\times10^{24}\big(h\nu\big)^4.
\label{app:eq:uv}
\end{equation}
The combined radiation field is given by $J_{\nu}^{\rm ISRF} = J_{\nu}^{\rm no UV} + J_{\nu}^{\rm UV}$ in units of $\rm erg \,s^{-1} \,cm^{-2} \,sr^{-1} \,Hz^{-1}$.

\subsection{Testing with a different ISRF}
\begin{figure}
\includegraphics[scale=0.51]{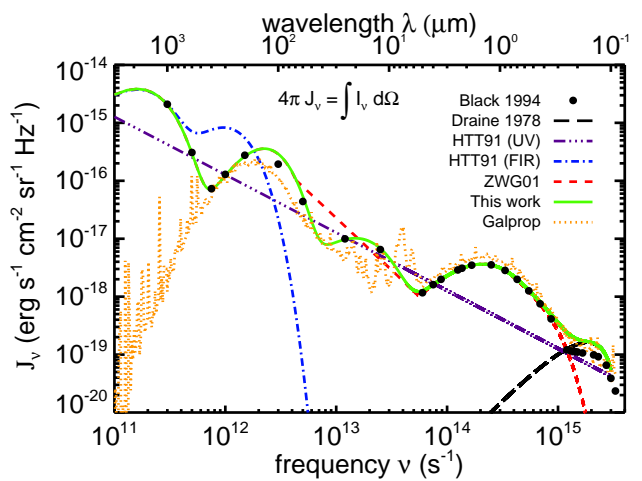}
\includegraphics[scale=0.51]{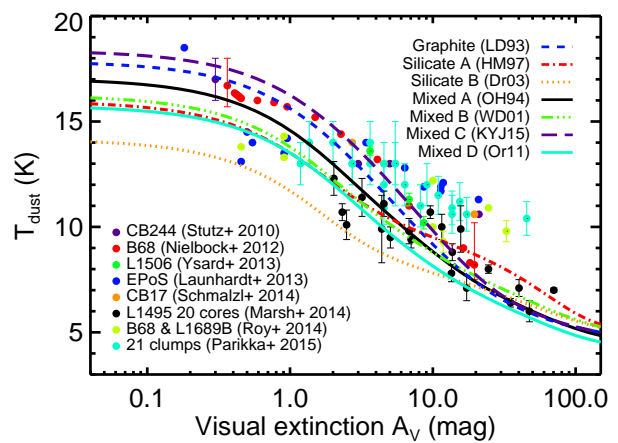}
\caption{In the same way as Figs.\,\ref{fig:eq1parameters} and \ref{fig:td1}, the ISRFs (top) and the dust temperature solutions (bottom) are plotted. Here, the \textit{Galprop} ISRF (yellow dotted line, top) is used for calculating the dust temperatures.}
\label{app:fig:galprop}
\end{figure}

In order to see how strong a dependence on the chosen ISRF there is, an online available, observationally constrained, ISRF is taken from \textit{Galprop} \citep{2007ARNPS..57..285S, 2012ApJ...750....3A} and used for the computations. The \textit{Galprop} code calculates and extrapolates the ISRF for every part of the Milky Way. For this exercise, the Galactic plane value at a distance of 8.5\,Kpc is adopted. In Fig.\,\ref{app:fig:galprop}, we see the results for this ISRF.

The dust temperatures are slightly higher at the edge, but lower at high $A_{\rm V}$. The first is due to a higher radiation flux at optical and UV wavelengths from the \textit{Galprop} ISRF, while the latter follows from the excluded flux at FIR and microwave wavelengths (notice the missing CMB part in the \textit{Galprop} ISRF).

We point out that a recent study by \cite{2016A&A...593L...5S} derives a factor 4 higher flux for the local MIR and FIR components of the ISRF of L1689B.

\section{The temperature impact of ices}
\label{app:sec:ices}
\begin{figure*}
\centering
\includegraphics[scale=0.51]{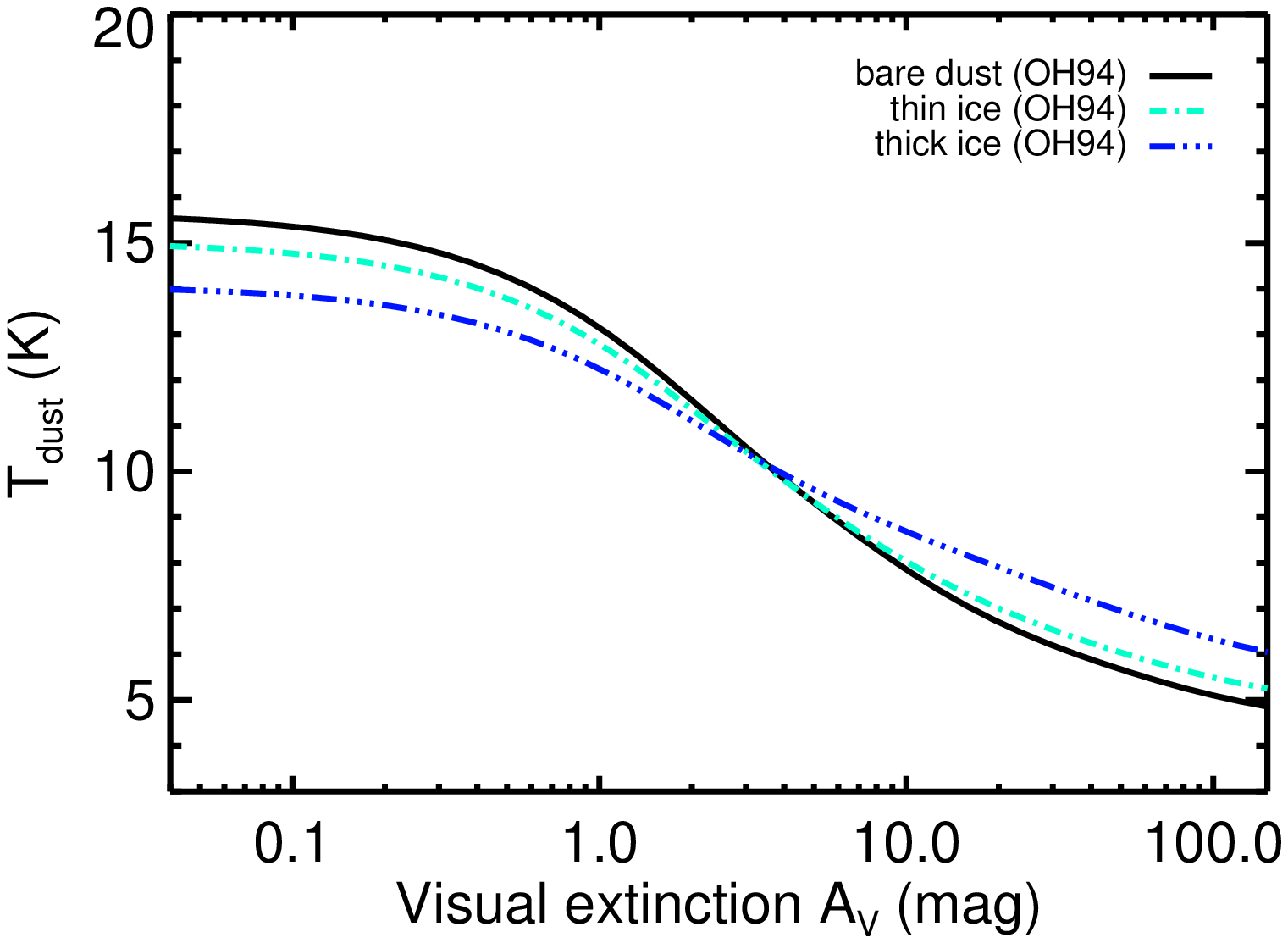} \\
\includegraphics[scale=0.51]{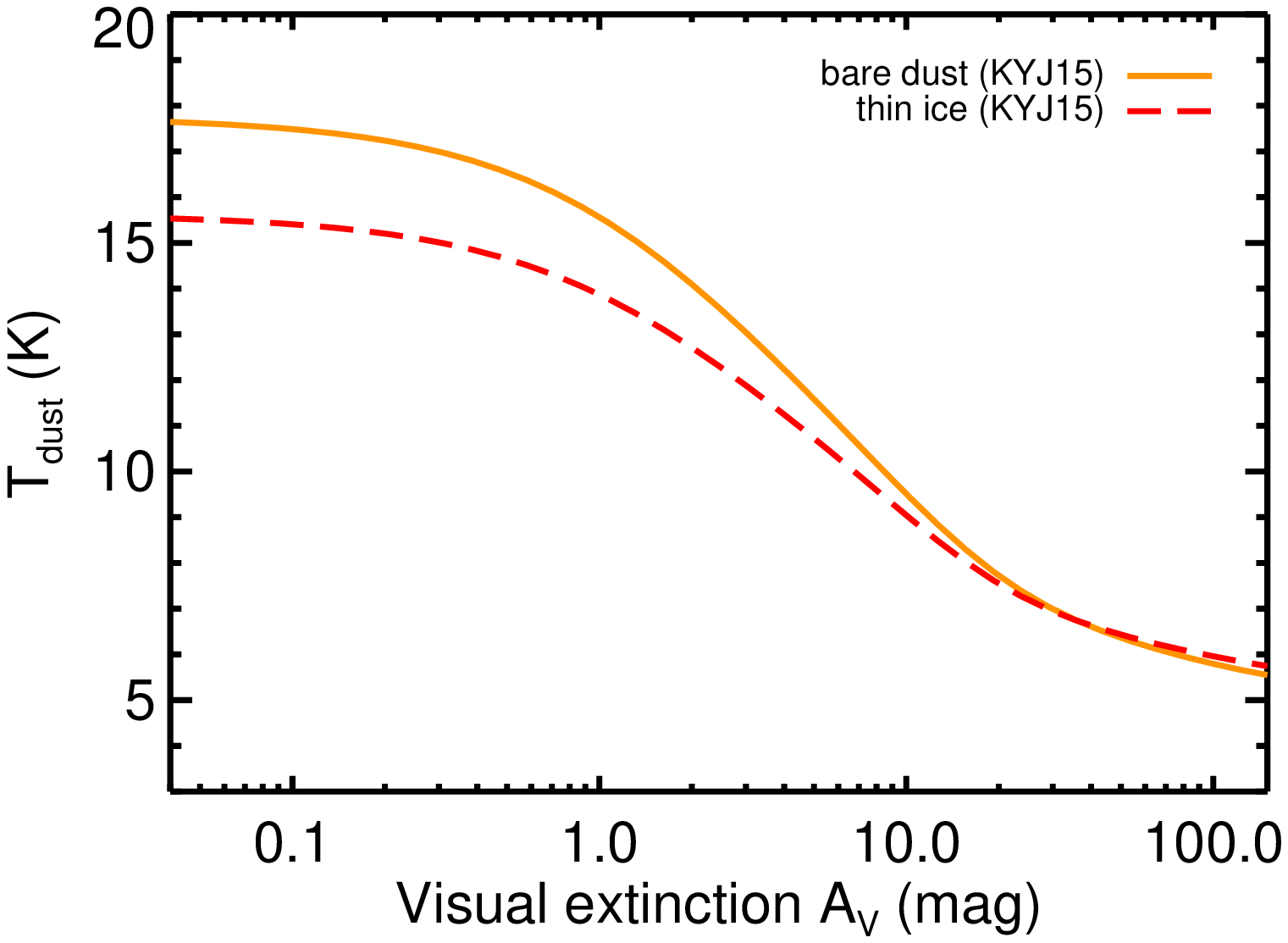}
\includegraphics[scale=0.51]{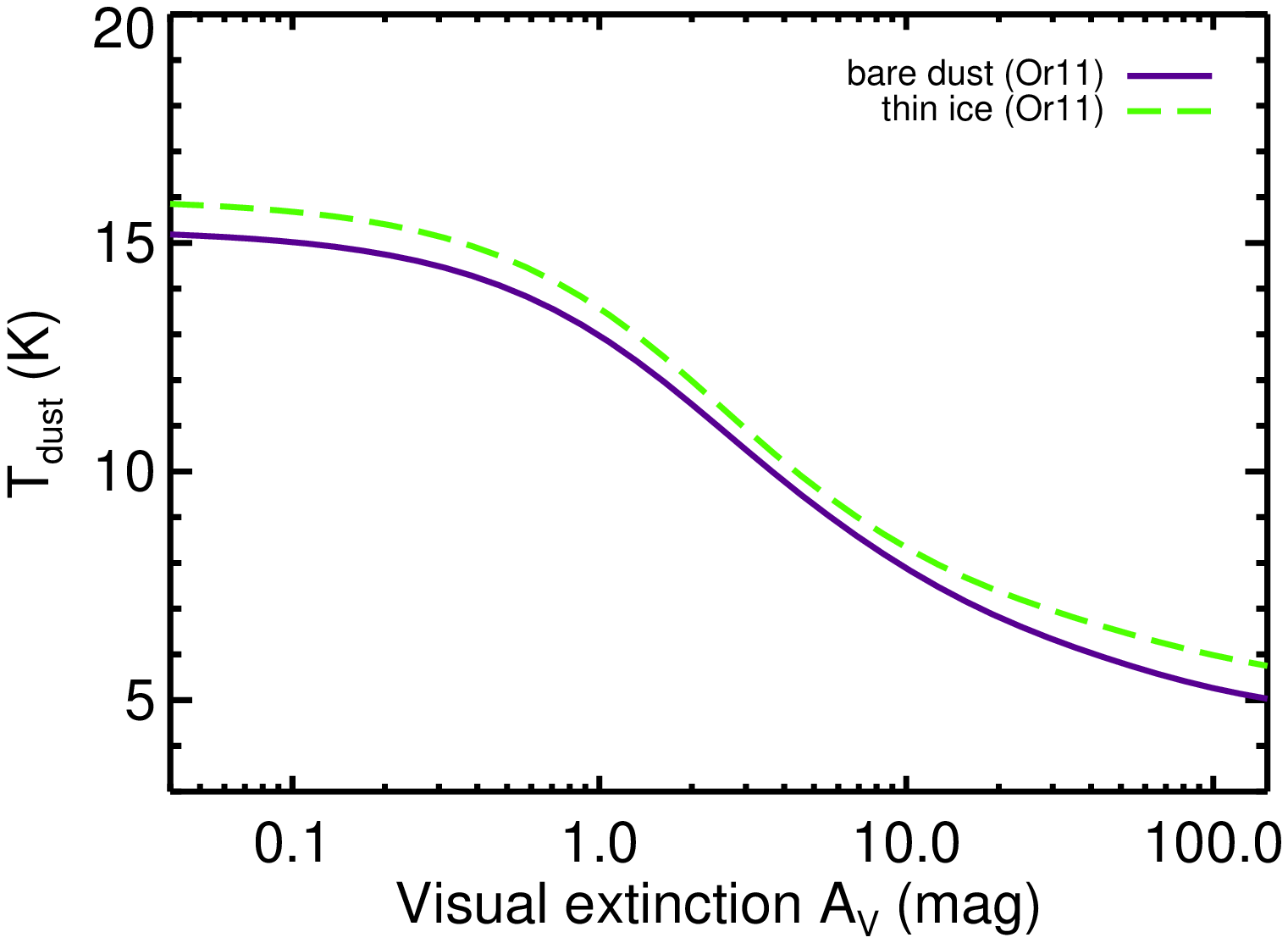}
\caption{$T_{\rm d}$ as a function of $A_{\rm V}$ with ices. Similar to the right panel of Fig.\,\ref{fig:td1}, the impact of ices on the dust temperature is displayed. Top panel shows the OH94 models, but coagulated for $10^5$\,yr, while the bottom panels show the KYJ15 (left) and the Or11 (right) models.}
\label{app:fig:ices}
\end{figure*}

Similar to what is shown in Fig.\,\ref{fig:td1}, right panel, the impact of ices on the dust temperature from two other studies, i.e., from Or11 and KYJ15, is displayed here. See Fig.\,\ref{app:fig:ices}. 
The sole exception is that the Or11 model does not reproduce the suggested trend at $A_{\rm V} < 0.8$\,mag, where the ice covered dust results in a higher $T_{\rm d}$. However, this is not relevant in a realistic case, because ices should not be present in this regime. While in his work Or11 does provide two other types of ice mixtures which yield more consistent results at low $A_{\rm V}$. It is advisable to note that, as stated before, the KYJ15 and the Or11 base ice models are not entirely separable from coagulation. The KYJ15 models include four big grain aggregates, while the Or11 models perform a minimum of 30\,Kyr coagulation, which is a relatively short timescale. Both of these models demonstrate the same trend as indicated by the OH94 models in Fig.\,\ref{fig:td1}.

\section{Comparing extinction curves}
\label{app:sec:extinction}
\begin{figure}
\includegraphics[scale=0.455]{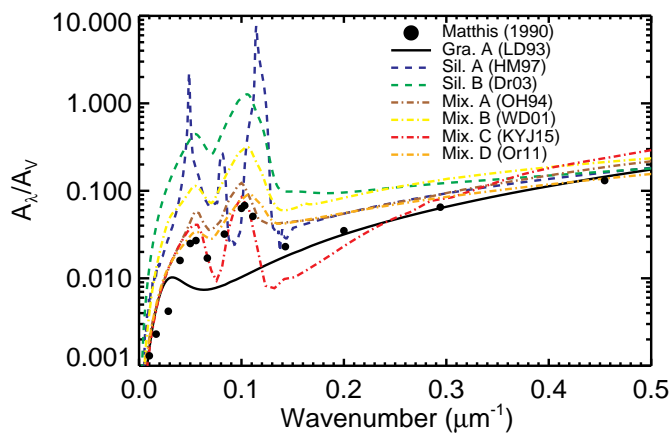}
\caption{Extinction coefficients $A_{\lambda}/A_{\rm V}$ as a function of wavenumber $1/\lambda$ ($\mu$m$^{-1}$). A comparison is made between \citet{1990ARA&A..28...37M}, black points, and self-consistent opacities, colored lines, used by \textsc{radmc-3d}.}
\label{app:fig:laszlo}
\end{figure}

The attenuation coefficient $\gamma_{\lambda} = A_{\lambda}/A_{\rm V}$ of the extinction curve from \cite{1990ARA&A..28...37M} and the self-consistent opacities of \textsc{radmc-3d} are compared, see Fig.\,\ref{app:fig:laszlo}. 

The \textsl{Mixed\,A} (OH94) model shows a very similar wavelength dependence in the range as the observed curve. For this model, the resulting temperature profiles from the semi-analytical and the \textsc{radmc-3d} methods are practically identical (Sect.\,\ref{sec:radmc3dresults}). The extinction curves resulting from pure silicate materials are very different than the observed extinction curve. At long wavelengths the differences are more complicated and highly wavelength dependent.

\section{Geometrical dependence}
\label{app:sec:geometry}
\begin{figure}
\includegraphics[scale=0.36]{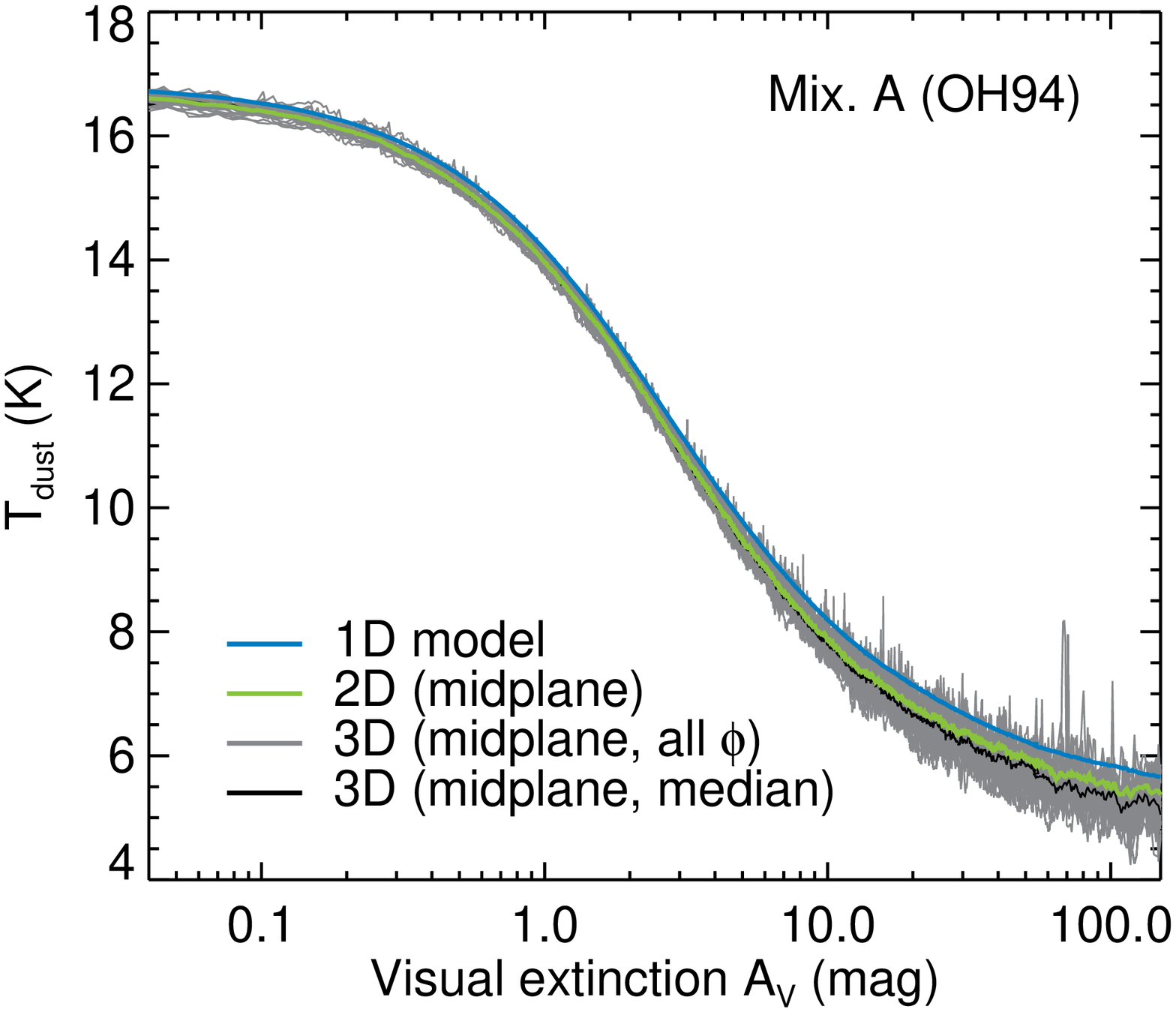}
\includegraphics[scale=0.36]{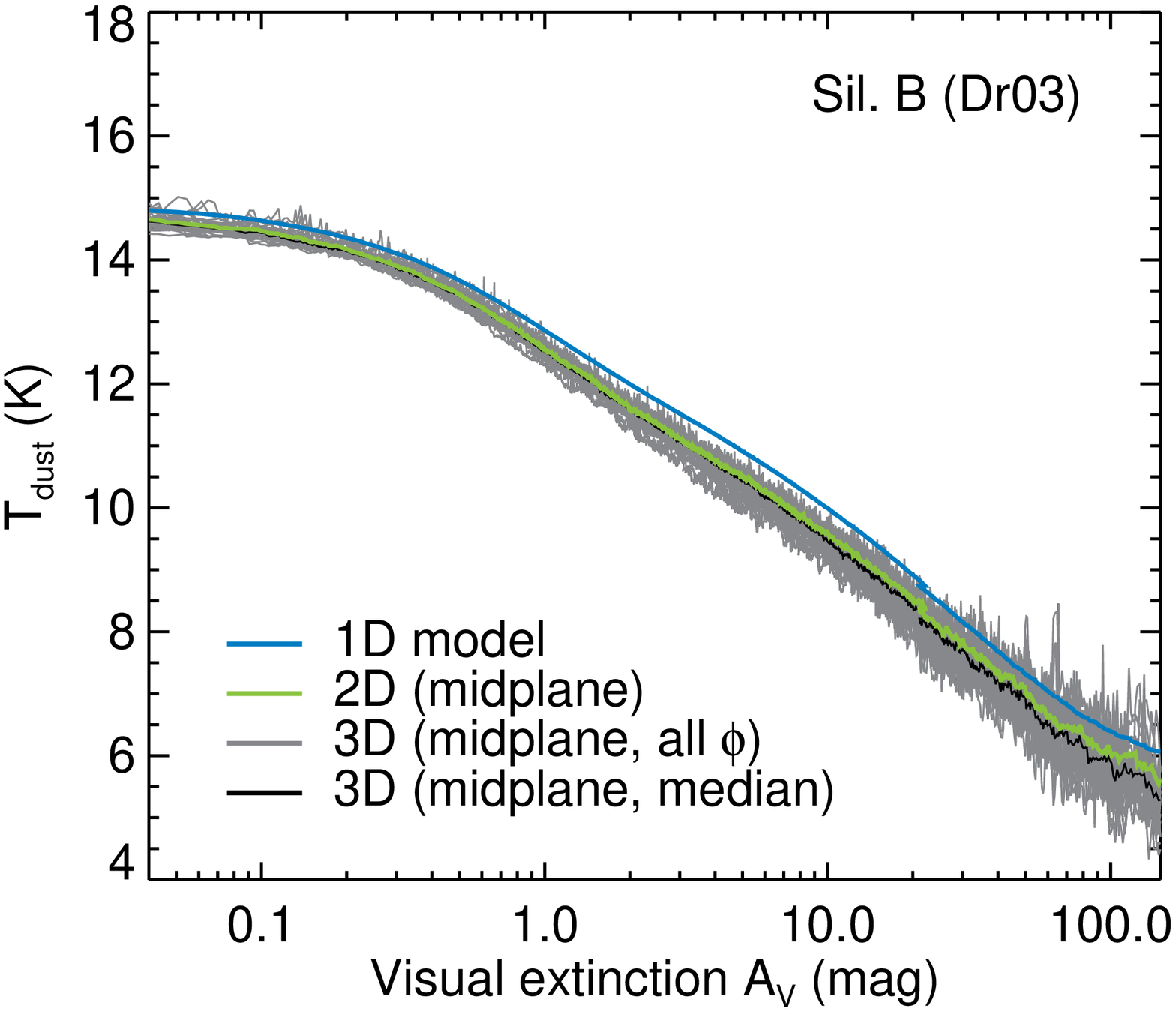}
\caption{The dust temperature in different geometries. The top panel shows the dimension impact for the \textsl{Mixed\,A} dust material type. The bottom panel shows this for the \textsl{Silicate\,B} dust material. Blue line shows the 1D model results, green line the 2D results, and black line the 3D results. The greyscales illustrate the variation from all the $\phi$ angles in the 3D model.}
\label{fig:dim123}
\end{figure}

The higher dimension models are set up similarly to the 1D model of \textsc{radmc-3d} (Sect.\,\ref{sec:radmc-3d}), a spherically symmetric cloud core with a power-law radial density profile (see Sect.\,\ref{sec:radmc3dcode}), but in spherical coordinates ($r, \theta, \phi$). In Fig.\,\ref{fig:dim123} the dimensional dependence is shown for two separate models: \textsl{Mixed\,A} (OH94) and \textsl{Silicate\,B} (Dr03). 

With higher dimensions, due to the increased number of grid cells, and without changing the number of photon packages, one loses precision. This results in a higher noise with increasing $A_{\rm V}$. One can particularly notice this from the 3D curves. Since there are the angles, $\theta$ and $\phi$, to be considered, for simplicity, the midplane value for $\theta$ at higher dimensional geometries are adopted. However, all $\phi$ angles for the 3D model is given in greyscales. The median radial temperature profile along the $\phi$ coordinate is drawn in black. Between all dimensions, and for both dust models, the difference is always less than 0.7\,K, that is, when considering the median value for the 3D model. There seems to be some geometrical dependence, but this dependence is small. The 1D model is consistently higher in temperature than the higher spatial dimension solutions. Due to the introduced noise, and the finer details of the code, it is hard to state if there is any systematic variation between the 2D and the 3D results.

\section{Analytical expressions from the literature}
\label{app:sec:literature}
There are several other studies that describe a method to calculate the dust temperature from first principles, assuming thermal equilibrium, and provide simple parametric expressions that can be used in astrophysical models. We report here solutions by three different groups in order to have a basis for comparison. The parametric expressions that will be discussed are acquired from \cite{1991ApJ...377..192H}, \cite{2001A&A...376..650Z}, and \cite{2011ApJ...735...15G}, which we had identified as HTT91, ZWG01, and GP11, respectively.

\subsubsection{The HTT91 expression}
The solution by \cite{1991ApJ...377..192H} assumes a one-sided slab geometry and the expression for the dust temperature is formulated as
\begin{eqnarray}
\nonumber
T_{\rm d}^{\rm HTT}\bigl(A_{\rm V},G_0\bigr) = \Bigl\{ 8.9\times10^{-11} \nu_0 G_0 e^{-1.8 A_{\rm V}} + 2.7^5 + 3.4\times10^{-2} 
\\ 
\times \left[ 0.42 - {\rm ln}(3.5\times10^{-2} \tau_{100} T_0) \right] \tau_{100} T_0^6\Bigr\}^{1/5} ~\rm K.~~
\label{eq:tdhol}
\end{eqnarray}
Here, $\nu_0$ represents the main absorbing frequency over the visual and UV wavelengths and is suggested as $\nu_0 = 3\times10^{15}$\,s$^{-1}$, $G_0$ is the UV flux in terms of the Habing field \citep{1968BAN....19..421H}, $\tau_{100}$ is the emission optical depth at 100\,$\mu$m, and $T_0$ is the equilibrium dust temperature at the cloud edge due to the unattenuated incident FUV field alone. Given the condition for $T_0$, this parameter equates to $T_0 = 12.17\,G_0^{1/5}$\,K. HTT91 assume that the incident FUV flux equals the outgoing flux of dust radiation at $T_0$, such that $\tau_{100} = 2.7\times10^3 G_0 T_0^{-5}$. Knowing $T_0$ fixes $\tau_{100}$ to a value of 0.001.

The HTT91 expression is a useful function which combines the heating by UV, CMB, and the re-processed IR. It is a function of $G_0$ and $A_{\rm V}$, but the $A_{\rm V}$ dependence is only for the attenuation of the incident UV field. The authors assume a fixed scaling of 1.8 for the UV attenuation with frequency, which is roughly a Planck averaged value over the extinction curve. The expression is scalable with the background UV field such that it can be applied to photodissociation regions. The assumed one-sided slab geometry will not affect the solution for the cloud edge, but will shape the transition region and center temperature, unless large enough column densities are considered. $Q_{\nu}$ is set to unity for every $\nu>\nu_0$, and to $\nu/\nu_0$ when this is not the case. This gives a linear scaling with frequency for the absorption efficiency. In reality, however, the extinction features from the type of grain material should make it highly non-linear. The HTT91 expression is often adopted and highly referenced in other studies.

\subsubsection{The ZWG01 expression}
The expression provided by \cite{2001A&A...376..650Z} is defined as
\begin{eqnarray}
T_{\rm d}^{\rm ZWG}\bigl(A_{\rm V}\bigr) = \left\{ T_{\rm d}\bigl({\rm \textsc{vnir}}\bigr)^{5.6} + T_{\rm d}\bigl({\rm \textsc{mir}}\bigr)^{5.6} + T_{\rm d}\bigl({\rm \textsc{fir}}\bigr)^{5.6} \right\}^{1/5.6} ~\rm K,
\label{eq:tdzuc}
\end{eqnarray}
where VNIR, MIR, and FIR stand for contributions from the visual plus near-infrared (NIR), mid-infrared, and far-infrared, respectively. The individual dust temperatures are given as 
\begin{eqnarray}
\nonumber
T_{\rm d}({\rm \textsc{vnir}}) \!\!\!\! &\simeq& \!\!\! 43\,A_{\rm V}^{-0.56} - 77\,A_{\rm V}^{-1.28},
\\ \nonumber
T_{\rm d}({\rm \textsc{mir}})  \!\!     &\simeq& \!\!\! 7.9\,A_{\rm V}^{-0.089} \left( 1.8 - 0.098\,A_{\rm V}^{0.5} + 7.9\times10^{-5}\,A_{\rm V}^{1.5} \right)^{1/5.6},
\\
T_{\rm d}({\rm \textsc{fir}})  \!\!     &\simeq& \!\!\! 6.2 - 0.0031\,A_{\rm V}.
\label{eq:tdzuc2}
\end{eqnarray}
The ZWG01 solution for the dust temperature is designed for the range $10\lesssim A_{\rm V} \lesssim400$\,mag. An additional expression is provided for $A_{\rm V} = 1-2$\,mag, which we will not consider hereafter because of the narrow range.

The ZWG01 expression is based on the observed dust temperature of L1544 at various $A_{\rm V}$. These authors solve the thermal balance using observed quantities and provide a parametric fit to the solution. The dust temperature expression is solely a function of $A_{\rm V}$. The thermal balance is solved without considering the UV field thereby purely concentrating on the visual and IR part of the spectrum. This is justified, because the focus is on regions with high $A_{\rm V}$ where UV light will be extincted. The ISRF spectrum is represented by a sum of five modified black bodies employing the data of \cite{1994ASPC...58..355B} and \cite{1983A&A...128..212M}, and a power-law for the MIR. The dust opacities are adopted from OH94 for the thin ice, 10$^5$\,yr coagulated case. ZWG01 do this by considering the opacities of a few peak frequencies and assume a power-law behavior rather than using the entire opacity table. They scale $Q_{\nu}$ as $(\nu/\nu_{\rm peak})^{\eta}$, where $\eta$ is obtained from the OH94 models.

The authors note that $T_{\rm d}^{\rm ZWG}$ can be scaled with the background radiation field by multiplying the dust temperature with a factor of the field intensity to the power 1/5.6 \citep{2001A&A...376..650Z, 2002A&A...394..275G}. This notion presumes, to an acceptable degree, that the whole spectral shape of the ISRF scales in the same manner, albeit the CMB should certainly not.

\subsubsection{The GP11 expression}
The expression by \cite{2011ApJ...735...15G} is presented as
\begin{eqnarray}
T_{\rm d}^{\rm GP}\bigl(A_{\rm V}\bigr) = 18.67 \,\textrm{-}\, 1.637\,A_{\rm V} \,\textrm{+}\, 0.07518\,A_{\rm V}^2 \,\textrm{-}\, 0.001492\,A_{\rm V}^3 ~\rm K,~
\label{eq:tdgar}
\end{eqnarray}
which is only valid in the range $0\leq A_{\rm V} \leq10$\,mag.

The GP11 formulation complements the low $A_{\rm V}$ regions and is designed to be used in combination with $T_{\rm d}^{\rm ZWG}$. The authors adopt a scaling of $Q_{\nu} \propto a\nu^2$ for the efficiency following \cite{2008ipid.book.....K}, and fix $a$, the grain radius, to 0.1\,$\mu$m. This gives a steep scaling to $Q_{\nu}$, which is only expected at long wavelengths. Their choice results in a high value of $Q_{\nu}$ below $\lambda\lesssim30\,\mu$m and a low value in the submillimeter. This results in a higher dust temperature at low $A_{\rm V}$. A critical note is that GP11 adopt the right-hand side of Eq.\,\ref{eq:thbalance} from ZWG01, which means that UV is not taken into account into the ISRF either. This is a rather important omission at low $A_{\rm V}$. GP11 also add a corrective value of 0.316 K to their solution to ensure continuity with ZWG01 at $A_{\rm V}=10$\,mag. For $D_{\nu}(A_{\rm V})$, see Eq.\,\ref{eq:thbalance}, the authors follow \cite{2006MNRAS.367.1757C} thereby using the tabulated values of \cite{1990ARA&A..28...37M}, times a factor 0.8, to evaluate $A_{\nu}/A_{\rm V}$. This is an improvement over fixing it to a single value as was the case in HTT91 and also over the three-part power-law method as was favored by ZWG01.

\subsubsection{Comparison between expressions}
\begin{figure}
\includegraphics[scale=0.51]{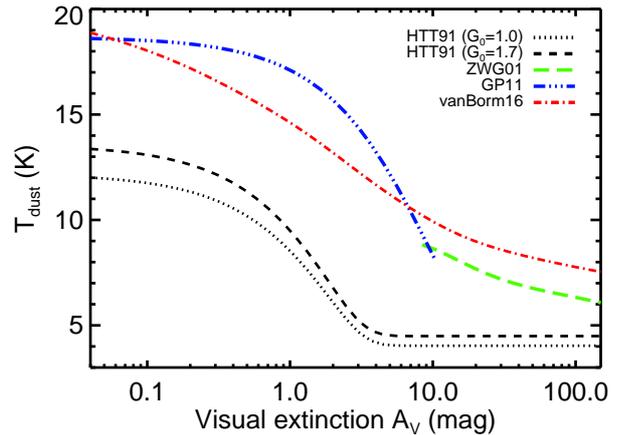}
\caption{Comparison between three analytical expressions for $T_{\rm d}$ as found in the literature.}
\label{fig:expressions}
\end{figure}

The three dust temperatures derived from the three expressions in the previous sections are presented in Fig.\,\ref{fig:expressions}. 
In this figure, we compare the three solutions for the ISRF as provided by the respective studies, which differ from each other.

It is clear that the differences between the models are quite large. The temperatures at cloud edge, which is the simplest condition because of no depth or geometry dependence, already vary between 12.2\,K and 18.9\,K. This disparity mainly arises from a different treatment of absorption efficiency ($Q_{\nu}$, see Fig.\,\ref{fig:eq1parameters}). At high optical depths, that is $A_{\rm V}=150$\,mag, the solutions of HTT91 and ZWG01 result in the dust temperatures of 4 and 6\,K. Here, the difference mostly comes from the different treatment of the FIR part of the spectrum. Among the three expressions, the dust temperature profile from the ZWG01 solution is more in line with the observations.

In Fig.\,\ref{fig:expressions}, we also show one extra curve, which is given by the red dot-dashed line. This is taken from \cite{2016arXiv160903900V}. Using a similar approach to ZWG01, the author derived analytically the dust temperature with a better treatment for the ISRF. It is worthwhile to note that this parametric expression reproduces the observed dust temperatures better than the ones from earlier studies. The full expression for $T_{\rm d}^{\rm \,vBorm}$ can be found in Chapter 3.2.H of that work.

\section{Scaling with background field}
\label{app:sec:uv}
\begin{figure}
\includegraphics[scale=0.51]{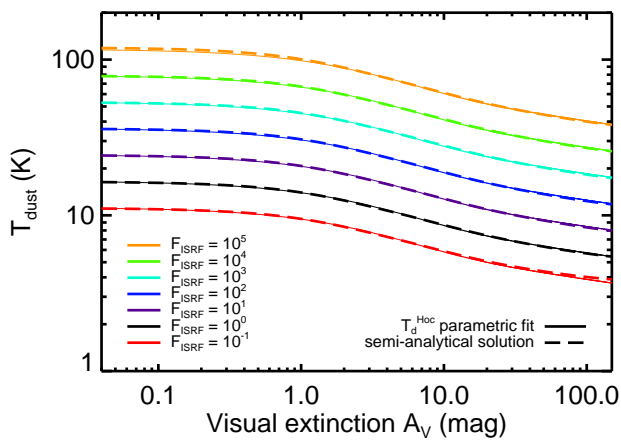}
\caption{The dust temperatures from semi-analytical solutions (dashed lines) and from parametric fits from Eq.\,\ref{eq:expression} (solid lines) for various ISRF strengths. The correspondence is noteworthy.}
\label{app:fig:multiG0}
\end{figure}

The thermal balance equation (Eq.\,\ref{eq:thbalance}) is also solved for various different radiation field strengths. The assumption is that $F_{\rm ISRF}$ (not including CMB) scales proportionally to the UV field, which is characterized by $\chi_{\rm uv}$. Here, the semi-analytical solutions are compared against the parametric fitting function for various radiation field strengths, i.e., $\chi_{\rm uv}$ = 0.1, 1, 10, $10^{2}$, $10^{3}$, $10^{4}$, $10^{5}$ (Eqs.\,\ref{eq:expression} and \ref{app:eq:expression}). The CMB part of the radiation field spectrum is not scaled with this factor for the semi-analytical calculations. The example given in Fig.\,\ref{app:fig:multiG0} is for mixed grains, bare dust. The parametric fits match the semi-analytical solutions excellently. We note that even at $A_{\rm V}=150$\,mag, heating by the ISRF dominates over the heating by the CMB.

\section{Additional expressions}
\label{app:sec:fits}
\begin{table*}
\centering
\caption{Three- and five-parameter fit coefficients}
\begin{tabular}{lccc}
\hline
\hline
Type & $\alpha$ & $\beta$ & $\gamma$ \\
\hline
Graphite						& 11.12 & 6.37 & 0.65 \\
Silicate grains 					& \multicolumn{3}{c}{not a good fit} \\
Mixed grains, bare dust					& 10.69 & 5.87 & 0.64 \\
$R_V = 3.1$, mixed bare dust				& 10.70 & 5.78 & 0.67 \\
Mixed grains, with ices					& 10.98 & 5.73 & 0.61 \\
$R_V = 3.1$, mixed with ices				& 10.96 & 5.60 & 0.64 \\
Mixed grains, with thick ices				& 11.23 & 5.56 & 0.56 \\
Observations, LOS corrected				& 11.68 & 5.80 & 0.58 \\
\hline
\end{tabular} ~~~
\begin{tabular}{lccccc}
\hline
\hline
Type & $\alpha$ & $\beta$ & $\gamma$ & $\delta$ & $\epsilon$ \\
\hline
Graphite						& 11.10 & 6.95 & 0.65 & -0.74 & 0.50 \\
Silicate grains						&  9.92 & 2.85 & 0.46 &  2.56 & 1.06 \\
Mixed grains, bare dust					& 10.41 & 5.12 & 0.62 &  1.06 & 1.40 \\
$R_V = 3.1$, mixed bare dust				& 10.57 & 4.97 & 0.67 &  1.04 & 0.94 \\
Mixed grains, with ices					& 10.56 & 4.76 & 0.58 &  1.41 & 1.52 \\
$R_V = 3.1$, mixed with ices				& 10.64 & 4.70 & 0.62 &  1.26 & 1.36 \\
Mixed grains, thick ices				& 10.75 & 4.14 & 0.52 &  1.96 & 1.29 \\
Observations, LOS corrected				& 11.44 & 4.66 & 0.60 &  2.47 & 0.68 \\
\hline
\end{tabular}
\label{app:tab:fits}
\end{table*}
Additional parametric expressions constructed from the fits to the semi-analytical solutions are given here. The fits are provided for every type of grain material (graphite, silicate, mixed), with and without ices, and for $R_V = 3.1$ (fiducial model is $R_V=5$). A slightly more accurate and an extended version of the expression in Eq.\,\ref{eq:expression} is provided here, i.e., 
\begin{eqnarray}
T_{\rm d}^{\rm Hoc} = \left[ \alpha + \beta\tanh\bigl(\gamma-X\bigr) + \delta\tanh\bigl(\epsilon-X\bigr)^3 \right] \chi_{\rm uv}^{1/5.9},
\label{app:eq:expression}
\end{eqnarray}
where $X = \log_{10}(A_{\rm V})$. The coefficients $\alpha$, $\beta$, $\gamma$, ($\delta$, $\epsilon$) are given in Table\,\ref{app:tab:fits} for the three-parameter and the five-parameter fits.

\begin{figure}
\includegraphics[scale=0.51]{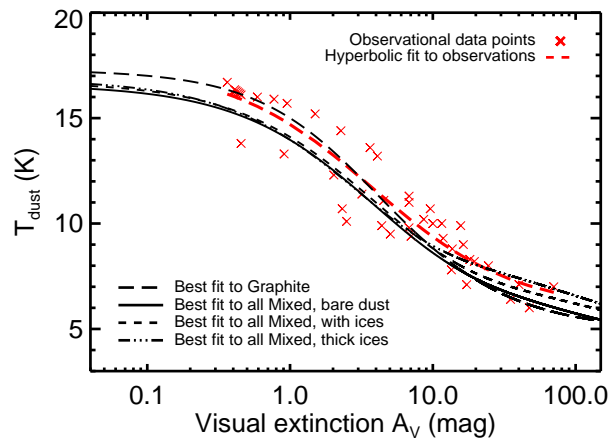}
\caption{Comparison between selected (graphite, mixed, and with ices) hyperbolic fits overlaid on top of observations (red crosses). Observational data only include LOS corrected points. Including ices into the mix brings the solutions nearer to the observed dust temperatures.}
\label{app:fig:fits}
\end{figure}

In Fig.\,\ref{app:fig:fits}, the difference between the various parametric fits, considering the 5-parameter fits, are highlighted and plotted on top of the observed dust temperatures. The observed $T_{\rm d}$'s, consisting of only LOS corrected data here, are fit with the hyperbolic function of Eq.\,\ref{app:eq:expression}, rather than the linear fit given in Sect.\,\ref{sec:obs}.

\section{Scaling with redshift}
In order to scale our expression with redshift $z$, the increase in CMB temperature has to be considered. The CMB temperature will rise according to the relation $T_{\rm CMB}(z) = T_{\rm CMB}^{z=0}\times(1+z)$. This will change the dust temperature solutions in the following way
\begin{equation}
T_{\rm d}^{\rm Hoc}(z) = \left[ \left(T_{\rm d}^{\rm Hoc}(z=0)\right)^{5.9} + \left(T_{\rm CMB}^{z=0}\right)^{5.9} \left( (1+z)^{5.9} - 1 \right) \right]^{1/5.9} \!\!\!\!\!\!\!\!,
\label{app:eq:redshift}
\end{equation}
see, for example, \cite{2013ApJ...766...13D}.

This simple prescription is not taking into account the changes in the UV field at high redshift due to different star formation rates and various other processes.

\end{document}